\documentclass[10.5pt,twocolumn,showpacs,showkeys,preprintnumbers,amsmath,amssymb,prb,superscriptaddress]{revtex4-2}
\usepackage[utf8]{inputenc}
\usepackage{amsmath}
\usepackage{amsfonts}
\usepackage{epsfig}
\usepackage{bm}
\usepackage{setspace}
\usepackage{booktabs} 
\usepackage{graphicx}
\usepackage{float}
\usepackage[section]{placeins}
\usepackage{siunitx}
\usepackage{multirow}
\usepackage{array}
\usepackage{placeins}
\usepackage{afterpage}

\newcolumntype{C}[1]{>{\centering\arraybackslash}m{#1}}
\usepackage[colorlinks=true, citecolor=red, linkcolor=blue, urlcolor=blue]{hyperref}
\begin{document}
\title{Unveiling AlSb as a Promising Zincblende Semiconductor for Visible-Light Shift-Current Generation}
\author{C\'esar Castillo-Quevedo}\email[email:]{cesar.cquevedo@academicos.udg.mx}
\affiliation{Departamento de Ciencias B\'asicas, Centro Universitario de la Ci\'enega, Universidad de Guadalajara,  Av. Universidad, Núm.1115, Col. Lindavista, C\'odigo Postal 47810, Ocotl\'an, Jalisco, Mexico,}
\author{Edgar Paredes-Sotelo}
\affiliation{Departamento de Investigaci\'on en Pol\'imeros y Materiales, Edificio 3G. Universidad de Sonora. Hermosillo, Sonora, M\'exico}
\author{Peter L. Rodr\'iguez-Kessler}
\affiliation{Centro de Investigaciones en \'Optica A.C., Loma del Bosque 115, Col. Lomas del Campestre, León, Guanajuato 37150, Mexico }
\author{Gerardo Mart\'inez-Guajardo}
\affiliation{Unidad Acad\'emica de Ciencias Qu\'imicas, \'Area de Ciencias de la Salud, Universidad Aut\'onoma de Zacatecas, km. 6 Carretera Zacatecas-Guadalajara S/N, Ejido La Escondida C.P. Zacatecas 98160, Zac, México}
\author{Jose Luis Cabellos}\email[email:]{jose.cabellos@uptapachula.edu.mx}
\affiliation{Universidad Polit\'ecnica de Tapachula, Carretera Tapachula a Puerto Madero km 24, San Benito, Puerto Madero C.P. 30830, Tapachula, Chiapas, M\'exico}
\date{\today}

\begin{abstract}
We use density functional theory to investigate the shift-current response in zincblende III–V (AlP, AlAs, AlSb, GaP, GaAs, InP, InAs, and InSb) and II–VI (ZnS, ZnSe, ZnTe, CdS, CdSe, and CdTe) semiconductors. Our main goal is to identify which material generates the largest shift-current under illumination and to examine the factors influencing this response. We find that aluminum-containing semiconductors, particularly AlSb, exhibit the highest shift-current responses, while CdSe shows the lowest. We analyze the contributions of specific band-to-band transitions to the shift-current in AlSb by selectively summing valence and conduction bands. Additionally, we calculate delocalization indices to investigate the electron delocalization, which correlate with the shift current. Hydrostatic pressure does not enhance the shift current in these materials. These findings have potential applications in optoelectronics and identify the most promising zincblende semiconductors for efficient shift-current generation under visible-light illumination.
\end{abstract}
\keywords{DFT, Shift current response, BPVE,   zincblende semiconductors, band-by-band, delocalization indices}
\maketitle

\section{Introduction}
Due to the declining availability of non--renewable energy resources, the efficient conversion of solar light into electricity is crucial for future clean-energy technologies, given that they contribute to greenhouse gas emissions reduction~\cite{GIELEN201938, LI20111785}.   Solar energy is an unlimited and cost-free energy resource, with approximately 1366 watts per square meter (W/m$^2$) reaching the top of the atmosphere~\cite {LI20111785, GUEYMARD20182}. This energy encompasses both light and heat, which are sources of electrical current. The primary focus is on the photovoltaic effect, which converts light into electric current~\cite{PhysRevLett.127.127402,PhysRevB.23.5590, TANG2023232785}  by illuminating a semiconductor p-n junction. Nonetheless, the theoretical efficiency limit for a single p--n junction is defined by the Shockley--Queisser (SQ) limit~\cite{10.1063/1.1736034,Qayoom_2023}. According to the SQ limit, the maximum theoretical efficiency of solar cells made from crystalline silicon p-n junction is {33.7\%}~\cite{ZANATTA2022100320,WANG2022756}.

To overcome the SQ limit,  alternative methods for photocurrent generation can achieve higher efficiencies. These methods include: the single-nanowire solar cells~\cite{Krogstrup2013},  the intermediate band solar cells~\cite{https://doi.org/10.1002/adma.200902388,https://doi.org/10.1002/pip.360,https://doi.org/10.1002/pip.3351}, hot--carrier cells~\cite{Li2017,10.1063/1.331124,10.1063/1.3489405,Qayoom_2023,C8TC04641G},   multiple--exciton generation~\cite{doi:10.1126/science.1209845,PhysRevB.108.125133} and particularly  the bulk photovoltaic effect (BPVE)~\cite{PRXEnergy.2.013006,Li2021,Zhang2022} or  the photogalvanic effect (PGE)~\cite{10.1063/5.0101513,PhysRevB.103.195203} which  offers a promising approach  to exceed  the SQ limit~\cite{Liang2023,Spanier2016,Pi2023}, thus compete with conventional solar cells~\cite{Tan2016, Cook2017,doi:10.1021/acs.nanolett.4c03944}. However, recently, a work pointed out that although the BPVE is not subject to the SQ limit~\cite{Sauer2023,Tan2016,10.1063/5.0101513,PhysRevB.104.235203}, the BPVE energy conversion eﬃciencies are orders of magnitude below the SQ limit of single--junction solar cells~\cite{PRXEnergy.2.013006}. Additionally, due to the limitations of the second-order response mechanism, the currents generated by BPVE may be smaller than those generated by conventional p-n junction~\cite{PhysRevB.100.245206,PhysRevMaterials.8.025001}.

The BPVE is a second-order nonlinear optical effect that converts light into a DC photocurrent when a homogeneous material lacking inversion symmetry is under illumination~\cite{PhysRevB.100.064301, DANG20222659,Liang2023,PhysRevLett.109.116601,PhysRevLett.127.127402,10.1063/5.0101513,Tan2016,PhysRevB.97.245143,sturman1992photovoltaic,Osterhoudt2019,https://doi.org/10.1002/adma.201505215}. This effect scales proportionally with the square of the incident electric field~\cite{PhysRevB.100.064301,Tan2016, Nakamura2017} and is characterized by a third-rank tensor, it also depends of the polarization of light. Furthermore, the BPVE  occurs as an ultrafast effect~\cite{PhysRevB.100.064301},  which makes it suitable for applications in ultrafast photon detection~\cite{PhysRevResearch.6.013123}.  In non-centrosymmetric homogeneous semiconductors, the primary contributions to the BPVE are the injection current and the shift current~\cite{PhysRevB.100.064301,PRXEnergy.2.013006, https://doi.org/10.1002/adom.202400651, doi:10.1073/pnas.2424294122}. Nevertheless, other previous studies indicate that the main contribution to BPVE is the shift current~\cite{Jin2024,PhysRevLett.109.116601, 10.1063/5.0101513,Sipe2000,PhysRevB.101.235448, PhysRevB.108.075413,PhysRevB.101.045104,doi:10.1021/acs.jpcc.4c03673, Qian2023,PhysRevLett.109.236601,PhysRevLett.110.057201,Tan2016,DANG20222659,Tan2016}. Specifically, the shift current arises due to a change in the electron position, while the injection current results from a change in group velocity of electrons during the interband transition~\cite{PhysRevX.10.041041, Sipe2000, PhysRevB.111.155402, Puente-Uriona2023,PhysRevB.97.245143}.  Importantly, illuminating semiconductors with circularly polarized light generates the injection current ~\cite {deJuan2017, Pi2023}. In contrast, the shift current can be generated by light regardless of its polarization~\cite{Pi2023, Sipe2000, PhysRevX.10.041041, Nakamura2017, PhysRevB.104.115402}. However, we point out that polar materials can induce shift currents with polarized and unpolarized light, whereas non-polar materials require polarization~\cite{Sauer2023, Tan2016}. Tiwari et al.~\cite{PhysRevB.101.235448} study the shift current in the $\alpha$-In$_2$Se$_3$ semiconductor, highlighting the impact of the polarization direction on the shift current spectrum.

We emphasize that for either linear or circular polarized light absorption in a semiconductor with photon energies above the band gap excites electrons from valence bands to conduction bands, generating shift current~\cite{Cook2017, C4EE03523B}. However, the shift current vanishes for photon energies below the band gap, indicating that the shift current is a resonant response.  It is also important to note that the rectification current is the unique second-order nonlinear response induced in semiconductors illuminated by laser fields with photon energies below the band gap~\cite{Nastos2006,https://doi.org/10.1002/adom.202403030}.  The shift current is intimately related to the quantum geometry of the electronic Bloch wave functions~\cite{PhysRevResearch.4.013164, PhysRevX.10.041041}. Furthermore,  the shift current depends on the Berry connections and the phase of the interband velocity matrix element. Those berry connections or  Berry vector potentials~\cite{RevModPhys.82.1959}  are closely related to Berry phases. Ultimately, the shift current is a quantum geometric response~\cite{PhysRevResearch.4.013164,Xiong_2021}.

The shift current have been theoretically studied as a microscopic mechanism for the  BPVE in a nonsymmetric material~\cite{PhysRevB.23.5590,PhysRevB.19.1548}. Furthermore, it was derived within the framework of Green’s functions~\cite{PetrKral_2000} and nonlinear optics response theory using a perturbative scheme in length or velocity gauge~\cite{PhysRevX.11.011001,Sipe2000, PhysRevB.23.5590, doi:10.1126/sciadv.1501524,PhysRevX.10.041041,Gao_2021,Ishizuka_2017,PhysRevLett.126.177403}. Young and Rappe were pioneers in comparing theoretical shift currents, calculated from first principles,
with  measured photocurrents in BaTiO$_3$ material~\cite{PhysRevLett.109.116601}.
Previous work presented a ﬁrst-principles algorithm based on maximally localized Wannier functions and
independent-particle approximation for computing the shift-current response~\cite{PhysRevB.97.245143}. Moreover, a recent
study of the shift and injection currents utilized numerical simulations based on mean-field theories to explore real-time
electron dynamics under laser pulse excitation,  the authors found that 
studies of many-body effects on the bulk photovoltaic effect should be carefully conducted~\cite{PhysRevB.109.195205}. 
Earlier work presents an expression for the many-body shift current from time-dependent perturbation theory from 
first principles using  Greens function times the screened Coulomb interaction W (GW approximation)
demonstrating significant  enhancement for the shift current ~\cite{doi:10.1073/pnas.1906938118}. Moreover,
to account for exciton effects, the GW approach plus Bethe Salpeter Equation (GW-BSE approach) was considered~\cite{lai2024bulkphotovoltaiceffectorigin}.  A recent GW-BSE calculation for shift currents in hexagonal boron nitride systems  attributed the enhancement of the shift current to excitonic-effects~\cite{PhysRevB.108.075413}.

The BPVE was first observed in BaTiO$_3$ material~\cite{DANG20222659}. Later studies investigated the photocurrent in a homogeneous, non--symmetric material under uniform and constant illumination, specifically in a ferroelectric system doped with iron and copper, particularly in the LiNbO$_3$ system \cite{10.1063/1.1655453}. We attribute the measured photocurrent to the shift current, which is recognized as a dominant effect in the BPVE.   Attention has primarily been focused on shift currents generated in ferroelectrics materials, which have attracted great interest as a candidate mechanism for achieving shift currents~\cite{Spanier2016,Grinberg2013,Nakamura2017,doi:10.1073/pnas.1802427116,PhysRevB.96.241203,Qian2023,https://doi.org/10.1002/adma.202301172,10.1063/5.0074371}. This includes ferroelectrics based on perovskite~\cite{Tan2016,PhysRevLett.109.236601,Kim2020,PhysRevLett.109.116601,PhysRevB.100.085102,PhysRevApplied.4.054004,PhysRevB.101.045104,PhysRevB.90.161409,Grinberg2013,D2RA06860E}, which have captured attention as materials for next-generation photovoltaic technologies. Ferroelectric materials exhibit a switchable polarization direction~\cite{Tan2016,https://doi.org/10.1002/adma.201505215}. In semiconductors, the difference in polarization between the valence and conduction bands leads to efficient generation of shift current~\cite{PhysRevB.96.075421}. Thus, the above suggests that ferroelectrics are candidates for materials capable of generating large shift-currents.

Many previous studies  have investigated the shift current in two-dimensional  materials~\cite{PhysRevMaterials.7.074001,PhysRevLett.119.067402,  doi:10.1021/acsomega.0c01319,Qian2023,PhysRevResearch.6.013123,Cook2017,PhysRevB.111.155402,doi:10.1073/pnas.2314775120,PhysRevLett.133.186801,Esteve-Paredes2025}. Notably, a particularly interesting study  is the published work where the shift current response of 326  2D semiconductors of the C2DB database have been calculated using  DFT~\cite{Sauer2023}.  The authors of this study found that the Janus compound {CrSTe} exhibits the highest efficiency for generating  shift current. Previous investigations have computed the shift current in single-element two-dimensional ferroelectrics, specifically phosphorene-like monolayers of arsenic (As), antimony (Sb), and bismuth (Bi). The authors reported a shift current of 2200  $\mu$A/V$^2$  for monolayer arsenic~\cite{Qian2023}.  In addition to  2D materials, special attention has been given to Weyl semimetals.  Recently, a work reported a large mid-infrared shift current in  Weyl semimetal TaAs~\cite{Osterhoudt2019}. Furthermore, Li et al. reported that shift current dominates second harmonic generation in TaAs~\cite{PhysRevB.97.085201}. Previous work has also explored the relationship between shift current and quantum geometry in Weyl materials~\cite{PhysRevX.10.041041}.

Although the shift current generation is more favorable in ferroelectric materials, previous experimental or theoretical studies reported shift currents in non-ferroelectric materials such as, GaAs, Te, ZnO, and HgS~\cite{10.1063/1.1436530,PhysRevB.97.245143, Sipe2000, PhysRevB.96.205201,PhysRevB.97.245143,https://doi.org/10.1002/adma.201505215, Fridkin1981}. In the context of perturbation theory and first-principles calculations, Nastos et al. presented the shift current for zincblende  GaAs and GaP semiconductors~\cite{Nastos2006}.  Their work demonstrates that the shift current dominates over the rectification current. Recently, another study~\cite{D4CP02478H} reported the shift current in both wurtzite and zincblende structures, such as AgI, GaAs, CdSe, CdTe, SiGe, ZnSe, and ZnTe semiconductors. The authors found that the shift current generated in wurtzite exceeds that generated in zincblende semiconductors, with the highest shift current computed in GaAs wurtzite, with the value of 31.8 $\mu$A/V$^2$.
 
In this article, we focus on studying the shift current response in zincblende semiconductors, specifically III-V compounds (AlP, AlAs, AlSb, GaP, GaAs, GaSb, InP, InAs, and InSb) and II-VI materials (ZnS, ZnSe, ZnTe, CdS, CdSe, and CdTe). The objective is to identify which of these semiconductors generates the highest shift current under uniform and constant illumination and to investigate the underlying mechanisms that enhance this shift current. All these compounds share the same symmetry operations, and the shape of the bands almost does not change for both valence and conduction bands near the $\Gamma$ point (similar electronic band structures). These characteristics of semiconductors allow us to study better the mechanism of generation of the shift current response on zincblende semiconductors and spintronics response~\cite{HARIHARAN2024111858}. 

Particularly, we identify that the zincblende AlSb semiconductor exhibits the highest shift current generation  in the visible spectrum. Moreover, our study investigates how the shift current is affected by the delocalization of the conduction and valence bands, as well as the chemical composition of the semiconductors. Our findings suggest that semiconductors containing aluminum are promising candidates for the efficient generation of shift current within or beyond the visible spectrum range. Materials containing antimony, particularly SbX (X=Al, Ga, In),  show potential for the efficient generation of shift current in the range of the visible light spectrum. This study on shift current generation demonstrates the potential of certain semiconductor materials to convert incident light into shift current.

 This paper is structured as follows.
In Section II, we review the zincblende structure.  
In section. III. We first
briefly review the general theory of the shift current response. In section IV, we describe the computational details. In Section IV, we present our results on the shift current. Finally, Section V is devoted to the conclusion. 

\section{Zincblende structure semiconductors}

\begin{figure}[ht!]
  \includegraphics[scale=0.2]{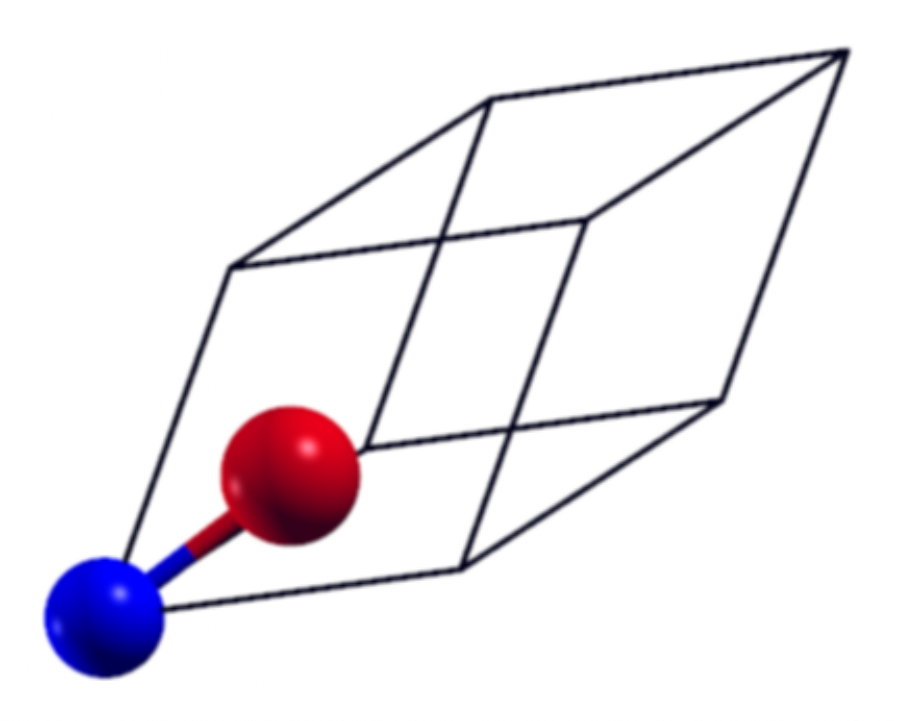}
  \caption{(Color online)  Primitive unit cell of gallium arsenide (GaAs) crystal in its zincblende structure with two atoms at the base. Blue and red spheres represent the Ga and As atoms, respectively.}
  \label{zincblendef}
\end{figure}

The zincblende structure is based on a face-centered cubic (FCC) lattice with two different types of atoms alternating in a specific pattern. It belongs to the space group 216 (F$\bar{4}3$m).  Figure~\ref{zincblendef} depicts the primitive unit cell of GaAs.  
The tetrahedral configuration of atoms within the zincblende structure is notably stable. In this arrangement, one type of atom, typically a cation, occupies the corners of the cube. At the same time, the other, usually an anion, is located at the face centers of the cube. For instance, in the gallium arsenide (GaAs) semiconductor, the cationic atoms, gallium, are positioned at the corners of the cube. In contrast, the anionic atoms, arsenic, are positioned at the face centers of the cube. Each As atom is located at the center of one of the cube's six faces. This specific arrangement results in the cubic symmetry characteristic of the zincblende structure~\cite{FEENSTRA1993251}. The point group associated with the zincblende structure is Td, which represents tetrahedral symmetry. The Td point group is nongyrotropic.  The zincblende structure has a cubic lattice, while the wurtzite structure has a hexagonal lattice. The atomic arrangement of the zincblende structure is similar to that of the wurtzite structure; the angle between adjacent tetrahedral units differs, measuring 60 for the zincblende and 0 for the wurtzite phase~\cite{10.1063/1.2787957}, or equivalently, in their dihedral conformation staggered and eclipsed for zincblende and wurtzite, respectively.  These structural differences are known as zincblende--wurtzite polytypism~\cite{PhysRevB.46.10086}. Further, the zincblende semiconductors  have lower ionicity compared to the wurtzite materials~\cite{10.1063/1.2787957}.

Furthermore, this structural symmetry streamlines the computation of electronic band structures, facilitating the prediction and design of electronic properties in zincblende photovoltaic devices~\cite{doi:10.1021/cm801911n,doi:10.1126/science.1069156}. 

\section{Theory}

The shift current response,  $J^a_{shift}(\omega)$, given in Equation~\ref{jota}, is related to a monochromatic time-dependent electric field, which has the form $E^b(t)=E^b(\omega)e^{i\omega t}+E^b(-\omega)e^{-i\omega t}$.
$J^a_{shift}(\omega)$ can be expressed in terms of the third-rank tensor or shift current tensor  $\sigma^{abc}(0;\omega,-\omega)$ . 
\begin{equation}
  \displaystyle
  J^a_{shift}(\omega)=2\sum_{bc}\sigma^{abc}(0;\omega,-\omega)E^b(\omega)E^c(-\omega).
  \label{jota}
\end{equation}

The shift current tensor is calculated within perturbation theory and under length gauge following Sipe et al.~\cite{Sipe2000,Nastos2006}
and T. Rangel~\cite{PhysRevLett.119.067402}. There, the shift current tensor is given by:
\begin{equation}\label{sigma1}
 \begin{split} 
   \sigma^{abc}(0;\omega,-\omega) & = - \frac{i \pi e^{3}}{2 \hbar^{2}} \int \frac{d^{3}\mathbf{k}}{8\pi^{3}}\sum_{nm}f_{nm}\\
   & (r^{b}_{mn} r^{c}_{nm;a} - r^{c}_{nm} r^{b}_{mn;a})
\delta(\omega_{mn} - \omega),
 \end{split}
\end{equation}

In Equation~\ref{sigma1},  $e$ and $\hbar$ are the electric charge and Planck’s constant, respectively.
$a, b, c$  letters are the axes x, y and z of an orthogonal cartesian coordinate system, and $\omega $ is the frequency of light. The sum is taken over all bands characterized by band indices $m$ and $n$, as well as crystal momentum $\mathbf{k}$, encompassing contributions from all electronic states within the material. The energy difference between bands $n$ and $m$ is given by $\omega_{mn}=\hbar\omega_{m}\mathbf(k) - \hbar\omega_n\mathbf(k)$;  the Fermi occupation at zero temperature is given by $f_{mn}-f_m-f_n$. The term $r^{a}_{mn}(\mathbf{k})$ represents the inter-band position matrix elements,  which are related to the Berry connection~\cite{Si_2025}. The  $r^{a}_{mn;b}(\mathbf{k})$ are the generalized derivatives given as $r^{a}_{nm;b}(\mathbf{k}) =\frac{\partial r^b_{nm}(\mathbf{k})}{\partial k^a}-i\big(A^a_{nn}(\mathbf{k}) -A^a_{nn}(\mathbf{k})\big)r^b_{nm}(\mathbf{k})$, with  $A_{nn}$ as the Berry connections of band $n$. Phenomenologically, the second-order response to the electric field reads $J_a=\sigma^{abc}E_bE_c$ the inversion operation converts $J_a \rightarrow -J_a$ and $E_{b,c}\rightarrow -E_{b,c}$, which implies $\sigma^{abc}=\sigma^{abc}$~\cite{sturman1992photovoltaic}. The terms shift BPVE and $j_{sh}$ were introduced in 1982 after a breakthrough in the understanding of the physics of the nondiagonal contribution to $j$. It was realized that an electron transition from the Bloch state is accompanied by y the shift $R_{nn}$ within a unit crystal cell~\cite{Sturman_2020}.
The shift current can be expressed as follows in the Equation~\ref{shift_current_response}:
\begin{equation}
  \begin{split} 
    \sigma^{abc}\left ( 0;\omega,-\omega  \right) & = -\frac{i\pi e^{3}}{2\hbar^{2}}\int \frac{dk}{8\pi^{3}} \sum_{n,m}f_{nm}\\
    & \left ( r_{mn}^{b} r_{nm;a}^{c} + r_{mn}^{c} r_{nm;a}^{b} \right) \\
    & \times \partial\left ( \omega _{mn}-\omega \right).\\
   \end{split} 
    \label{shift_current_response}
    \end{equation}
Here, $\sigma^{abc}\left(0;\omega,-\omega\right)$ describes the shift current  when a monochromatic laser beam is incident on a semiconductor.
\twocolumngrid

\section{Computational details}
We conducted geometry optimization and calculated electronic and optical properties, particularly the band structures and wavefunctions,  using density functional theory as implemented in the freely available ABINIT software package~\cite{Gonze2016}. 
To determine the total energy and equilibrium lattice constant, we performed geometry optimizations using the Broyden--Fletcher--Goldfarb-Shanno algorithm (BFGS)~\cite{HEAD1985264}. The atomic positions were relaxed until the maximum Hellmann-Feynman force per atom was less than  20 meV/\AA~. The self-consistent field (SCF) cycle was stopped once the total energy difference was less than 10$^{-9}$ eV.   For all atoms, we employ the Hartwigsen-Goedecker-Hutter (HGH) pseudopotential type~\cite{PhysRevB.58.3641}.,  which is a norm-conserving pseudopotential and includes multiple projectors and semi-core states. The HGH pseudopotentials use
analytical functions and incorporates scalar relativistic effects~\cite{Berry2024}.
These pseudopotentials do not include non-linear core corrections and can be used to perform mGGA
calculations~\cite{doi:10.1126/science.aad3000} with the
Tran-Blaha 09 XC functional~\cite{Tran2009}.  The HGH pseudopotentials have been employed in previous works for study the electrolytes in Zn batteries~\cite{Yang2023}, catalysis~\cite{doi:10.1126/science.adf6984}, the calculation of shift-currents~\cite{https://doi.org/10.1002/adom.202400651} and non-linear optical properties~\cite{PhysRevB.84.195326,PhysRevB.80.155205,PhysRevB.80.245204}.

For structure optimization, we employ the functional Perdew-Burke-Ernzerhof generalized gradient approximation (GGA) revised for solids known as PBEsol~\cite{PhysRevLett.100.136406}.  PBEsol differs from PBE~\cite{PhysRevLett.77.3865} only in two parameters~\cite{Zhang_2018} and belongs to the GGA level. LDA~\cite{PhysRev.140.A1133} and PBE functionals lead to under- and overestimations in computed lattice parameters, respectively~\cite{Yuk2024}. PBEsol yields reasonable lattice constant~\cite{PhysRevB.100.045135} and puts the lattice constant between those of LDA and PBE~\cite{PhysRevB.85.014111}. Particularly, PBEsol lattice constants are systematically lower and better than PBE by 1\%-2\%,~\cite{Zhang_2018}.

On the other hand, we computed the electronic and optical properties using the Tran--Blaha 09 XC exchange--correlation
functional~\cite{Tran2009}, also known as the modified Becke--Johnson~\cite{PhysRevB.83.195134} or simplicity mBJ. Since the mBJ is a potential functional, it is not applicable for calculating forces. Consequently, it is not suitable for optimizing geometries~\cite{PhysRevB.83.195134, D0TC05964A}. Therefore, we combine the optimized structure obtained from the PBEsol method for calculating the electronic structure and optical properties using the mBJ approach. Previous studies have shown that the mBJ method is highly accurate for calculating the band gap in semiconductors. Furthermore, other research employed the mBJ approach to compute the electronic and optical properties of chalcogenide--based zinc and cadmium monolayer semiconductors\cite{SAFARI2017663}, demonstrating good performance.

We expanded the wavefunctions using a plane-wave basis set with a kinetic cutoff energy of 40 Ha. To sample the irreducible Brillouin zone (IBZ), we employed the Monkhorst-Pack scheme~\cite{PhysRevB.13.5188}. The integration of the IBZ was performed using the tetrahedron method~\cite{ma13194300,PhysRevB.80.155205,PhysRevB.80.245204,PhysRevB.85.165324}.  We utilized a 40 x 40 x 40 mesh of $ k$-points to sample the IBZ, which was used for shift current calculations and electronic properties. Furthermore, we take into account the spin-coupling effect for all calculations as it influences the optical properties of these types of semiconductor materials~\cite{FENG20092103,PhysRevB.60.8610}. The indirect transitions were neglected~\cite{Okoye_2003}, along with the local field and electron-hole effects~\cite{PhysRevB.71.195209}. The inclusion of these effects is beyond the scope of this study. The shift current response, which depends on the momentum matrix elements and energy eigenvalues, was computed employing the in--house shift-current--response code developed by E. Paredes-Sotelo and J.L. Cabellos, based on TINIBA code~\cite{doi:10.1021/acs.nanolett.4c03880,PhysRevMaterials.6.034006,PhysRevMaterials.8.116203}. The calculation of the delocalization indices (DIs)  in zincblende periodic semiconductors studied here was carried out using both the critic2 code~\cite{OTERODELAROZA2009157,OTERODELAROZA20141007} and Quantum ESPRESSO code~\cite{Giannozzi_2017}. The delocalization indices provide a direct quantitative measure of interatomic electron delocalization~\cite{doi:10.1021/acs.jctc.8b00549}, and it could be correlated to the intensity of the generated shift current.  The  DIs calculation is based on the use of maximally--localized Wannier functions introduced by Marzari and Vanderbilt~\cite{RevModPhys.84.1419,PhysRevB.56.12847} and were obtained using the wannier90 code~\cite{MOSTOFI2008685}  with random starting projections.

\section{Results and discussion}
\subsection{Structural properties}
\begin{table}[htb!] 
\centering
\renewcommand{\arraystretch}{1.2} 
\begin{tabular}{
>{\raggedright\arraybackslash}p{1.5cm} 
>{\centering\arraybackslash}p{2cm} 
>{\centering\arraybackslash}p{2.2cm} 
>{\centering\arraybackslash}p{1.5cm}
}
\toprule
Material & Calculated a$_0$~(\AA) & Exp.~\cite{madelung2004semiconductors, tamargo2002ii} a$_0$~(\AA) & Deviation (\%) \\
\midrule
AlAs  &  5.6816& 5.6600 &  0.38  \\
AlP   &  5.4713& 5.4635 &  0.14  \\
AlSb  &  6.1691& 6.1355 &  0.55  \\
GaAs  &  5.6836& 5.6532 &  0.89  \\
GaP   & 5.4745 & 5.4505 &  0.44  \\
GaSb  & 6.1281 & 6.0959 &  1.00  \\
InAs  & 6.1243 & 6.0583 &  1.09  \\
InP   & 5.9320 & 5.8687 &  0.92  \\
InSb  & 6.5414 & 6.4793 &  0.96  \\
ZnS   &  5.3615& 5.4060 & -0.82  \\
ZnSe  &  5.6409& 5.6680 &-0.48  \\
ZnTe  &  6.0712& 6.1030 &-0.52  \\
CdS   & 5.8366 & 5.8350 &  0.03  \\
CdSe  & 6.0936 & 6.0500 & 0.72  \\
CdTe  &  6.4965& 6.4780 &  0.29  \\
\bottomrule
\end{tabular}
\caption{
  Calculated values at T=0 K and collected experimental values of lattice parameters (a$_0$) in~\AA~are listed for III-V and II-VI zincblende semiconductors.}   
\label{tab:lattice_parameters}
\end{table}
The computed equilibrium lattice constants  (in~\AA)~for  the II--VI and III--V zincblende semiconductors studied in this work are presented in Table~\ref{tab:lattice_parameters}.  The experimental reference values were collected from different published works. Moreover, the table includes the relative deviation (in percentage) of the computed equilibrium lattice constant versus experimental reference values. The DFT calculations describe ground-state properties at absolutezero (0 K)~\cite{PhysRevB.106.085137, D1CE00453K} ignoring usually the
thermal  effects~\cite{https://doi.org/10.1002/wcms.1294}.   The temperature induce some inconsistency between the theoretical and experimental results~\cite{molecules26133953,molecules26185710,molecules29143374}.  For instance, The lattice constants of many materials would increase a bit from 0 K to finite temperature, which usually results in slight change in the spatial distribution of electronic charge density~\cite{Wang2022}. The incorporation of temperature effects into elastic constant predictions in calculations accounts for approximately 30\% of the disagreement observed compared to elastic constants at 0 K~\cite{Wang2022}. We consider that the inclusion of temperature is beyond of this study.

The PBEsol functional provides an accurate reproduction of experimental lattice constants, with a relative deviation of less than 1.1\% for II--VI and III--V zincblende semiconductors, as displayed in Table~\ref{tab:lattice_parameters}.  The slightest deviation between the computed lattice constants and the experimental reference values is found in CdS, which belongs to the group of II--IV zincblende semiconductors.  The zinc chalcogenides (ZnS, ZnSe, and ZnTe) subestimated the lattice constants with relative deviations ranging from -0.52 to -0.85 percent. In contrast, the rest of the semiconductors overestimated the lattice constant. The possible reasons for those large underestimations of relative deviations in the zinc based group II-VI semiconductors. The lattice subestimations in zinc chalcogenides may be related to their ionic character since it showed that the lattice constant is related to ionic radii and electronegativity of the constituting ions of the material~\cite{BRIK20111256}. The electronegativity of the chalcogen atoms S, Se, and Te is S=2.58, Se=2.55, and Te=2.10, respectively. According to Table~\ref{tab:lattice_parameters}, the zinc chalcogenide ZnS possesses the largest relative deviation.    The cadmium chalcogenides present relative deviation in the range of 0.03 and  0.72\%  with CdS showing the smallest deviation; consequently, there is no clear trend between the zinc and cadmium chalcogenides. We have chosen to use the PBEsol exchange-correlation functional as it has been shown to accurately reproduce the experimental lattice parameters~\cite{PhysRevB.99.085207,PhysRevB.92.144308}.
\subsection{Bandstructure}
\begin{figure*}[htb!]
  \includegraphics[scale=0.80]{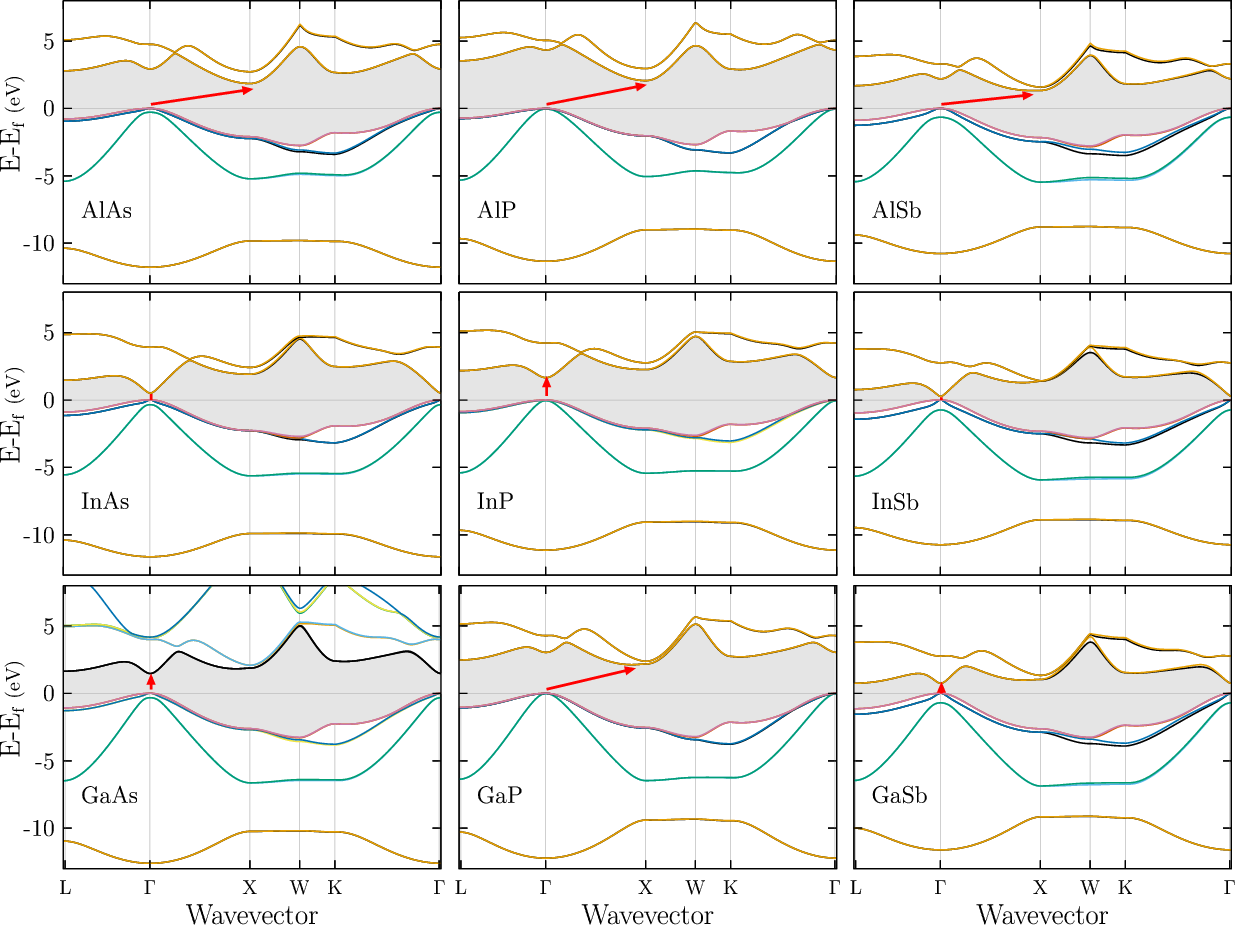}
  \caption{(Color online) The calculated band structure, along with the high--symmetry directions of the Brillouin zone for group III--V semiconductors, is presented here. The shaded region represents the forbidden band, while the red arrows indicate the band gap. This band structure is displayed without applying the scissors correction.}
  \label{bands1}
\end{figure*}

In Figure~\ref{bands1}, we show the results of our calculations for the electronic band structures of nine zincblende III--V semiconductors, using HGH pseudopotentials and taking into account the spin--orbit effect.  In Appendix~\ref{appendix:a}, Figure~\ref{bands2} depicts the electronic band structure for the zinc chalcogenides and cadmium chalcogenides studied in this work. The band structure of the group III--V bulk semiconductors has been reported previously~\cite{PhysRevB.82.205212, PhysRevB.87.235203,10.1063/1.1368156,PhysRevB.59.5536}, as well as the band structure of the zinc chalcogenides and cadmium chalcogenides~\cite{PhysRevB.50.10780,SAFARI2017663}. Due to the structural similarities of the semiconductors studied in this work, many of the high--symmetry points in the Brillouin zones are related to each other. Consequently, the band structures shown in Figure~\ref{bands1}  and Figure~\ref{bands2} are nearly identical in shape. However, we obtained different values of the band gap and spin--orbit splitting.

It is particularly interesting to explore the underlying mechanism of the enhancement of shift current and its correlation with band structure in group II--V and III--V semiconductors. Previous works in ferroelectric SnTe monolayers suggest that the optical transitions between the frontier bands near the valleys have the most significant effect on shift currents~\cite{Jin2024}. Additionally, earlier research suggests that a large density of states is essential to achieve a large shift current~\cite{Jin2024}.
In this study, we employed the primitive unit cell calculated at the PBEsol level of theory to compute all electronic and optical properties.  The band structure and optical calculations were conducted using the mBJ method. Employing a conventional unit cell to compute the band structure results in a folded band structure that may lose information; for instance, it is harder to visualize the density of states~\cite{PhysRevLett.104.216401, ma13194300}. Also, we considered the high--symmetric directions in the Brillouin Zone,  starting from the k--point L with coordinates (0.0,0.0,0.5), followed by the $\Gamma$  (0,0,0), X point (0,0.5,0.5), W point (0.25,0.5,0.75), K point (0.375,0.375,0.75), and  $\Gamma$ point (0,0,0), all expressed in units of 2$\pi$/a, 2$\pi$/b, 2$\pi$/c, where a, b, c,  are the lattice constants of the semiconductor.  Our calculated  SO--splitting and band gap values are compared with experimental data. For most zincblende semiconductors, the agreement is excellent. For instance, the computed SO--splitting value for GaSb (0.69 eV), ZnTe (0.92 eV), and AlAs (0.28 eV) are in excellent agreement with the experimental data as shown in Table~\ref{table_so2}. For those three systems, the difference between theory and experiment is less than 20 meV. However, there are several noticeable cases in which the difference is much larger. For instance, in the ZnS semiconductor, the calculated value is 0.01 eV, whereas the value from experimental data is 0.06 eV, and the difference between theory and experiment is 50 meV.  One cause of the discrepancy could be  the p-d hybridization at the  valence-band maximum  is overestimated, which may lead to smaller calculated SO--splitting~\cite{PhysRevB.70.035212,PhysRevB.37.8958}. The SO--effect is related to the atomic number Z; It is an important parameter for the determination of optical transitions in semiconductors~\cite{PhysRevB.70.035212}; Moreover, the SO--effect in InSe semiconductor allows optical transitions to couple with in-plane polarized light~\cite{PhysRevB.96.195428}. 
It is essential to consider the spin--orbit interaction because it removes the degeneracy of the valence band maximum~\cite{PhysRevB.88.045412}, resulting in six individual bands~\cite{10.1063/1.123461}.  

We now discuss details of the band structure shown in Figure~\ref{bands1} and the band gaps shown in Table~\ref{tab:lattice_parameters}. The Fermi level is shifted to 0 eV, and  the calculated band structures of all semiconductors are presented within the energy range of -12  to  8 eV, as displayed in Figure~\ref{bands1}.  The zincblende semiconductors usually have a valence band maximum located at the center of the Brillouin zone~\cite{Hutchings:92} ($\Gamma$ point), as we can see in Figure~\ref{bands1}. In the absence of the spin-orbit interaction,   the maximum valence band consists of three bands: a heavy--hole band and a light--hole band with a spin--orbit split--off band separated by the spin--orbit interaction energy (SO-splitting), which means that the valence band maximum at $\Gamma$ point degenerates for all the semiconductors considered in this study. It is now well established that both the three highest valence bands and the lowest conduction band of III--V and II--VI zincblende semiconductors are nonparabolic, nonspherical, and spin split~\cite{PhysRevB.76.125203}.
\begin{table}[htb!]
    \centering
    \renewcommand{\arraystretch}{1.0}
    \setlength{\tabcolsep}{2pt}
    \begin{tabular}{lllc}
         \toprule
        \textbf{Solids} & \hspace{1mm} \textbf{SO splitting } & \hspace{1mm} \textbf{Band Gap} & \textbf{Nature} \\
        & \hspace{2.6mm} Calc. / Expt. & \hspace{2.6mm} Calc. / Expt. & \\ 
         \midrule
        AlAs  & \hspace{0mm} 0.28 / 0.30\textsuperscript{a} & \hspace{0mm} 1.87 / 2.15\textsuperscript{b} & I \\
        AlP   & \hspace{0mm} 0.05 / 0.05\textsuperscript{a} & \hspace{0mm} 2.09 / 2.50\textsuperscript{b} & I \\
        AlSb  & \hspace{0mm} 0.65 / 0.75\textsuperscript{a,c,d} & \hspace{0mm} 1.35 / 1.61\textsuperscript{b} & I \\
        GaAs  & \hspace{0mm} 0.32 / 0.35\textsuperscript{a,d,e,f} & \hspace{0mm} 1.20 / 1.519\textsuperscript{m} & D \\
        GaP   & \hspace{0mm} 0.03 / 0.08\textsuperscript{a} & \hspace{0mm} 2.15 / 2.27\textsuperscript{b,g,h} & I \\
        GaSb  & \hspace{0mm} 0.69 / 0.70\textsuperscript{a,c,e} & \hspace{0mm} 0.50 / 0.75\textsuperscript{b,h} & D \\
        InAs  & \hspace{0mm} 0.34 / 0.35\textsuperscript{a,d} & \hspace{0mm} 0.30 / 0.35\textsuperscript{b} & D \\
        InP   & \hspace{0mm} 0.06 / 0.12\textsuperscript{a,e} & \hspace{0mm} 1.46 / 1.34\textsuperscript{b} & D \\
        InSb  & \hspace{0mm} 0.73 / 0.79\textsuperscript{a,e} & \hspace{0mm} 0.05 / 0.23\textsuperscript{b} & D \\
        CdS   & \hspace{0mm} 0.01 / 0.05\textsuperscript{a} & \hspace{0mm} 3.03 / 2.58\textsuperscript{i} & D \\
        CdSe  & \hspace{0mm} 0.40 / 0.84\textsuperscript{a,j} & \hspace{0mm} 2.03 / 1.77\textsuperscript{i} & D \\
        CdTe  & \hspace{0mm} 0.89 / 0.95\textsuperscript{a,c,j} & \hspace{0mm} 1.63 / 1.50\textsuperscript{i,k} & D \\
        ZnS   & \hspace{0mm} 0.01 / 0.06\textsuperscript{a} & \hspace{0mm} 4.30 / 3.74\textsuperscript{i} & I \\
        ZnSe  & \hspace{0mm} 0.42 / 0.40\textsuperscript{a,j} & \hspace{0mm} 2.93 / 2.82\textsuperscript{i} & D \\
        ZnTe  & \hspace{0mm} 0.92 / 0.91\textsuperscript{a,c,j} & \hspace{0mm} 2.29 / 2.39\textsuperscript{i,l} & D \\
        \bottomrule
    \end{tabular}
    \vspace{0mm}
    \footnotesize{
    \textsuperscript{a}~\cite{PhysRevB.70.035212}
    \textsuperscript{b}~\cite{madelung2004semiconductors}
    \textsuperscript{c}~\cite{10.1063/1.1607516}
    \textsuperscript{d}~\cite{PhysRevLett.11.541,Al-Douri2013}
    \textsuperscript{e}~\cite{PhysRevB.38.1806}
    \textsuperscript{f}~\cite{10.1063/1.2085170}
    \textsuperscript{g}~\cite{https://doi.org/10.1002/pssr.202300489}
    \textsuperscript{h}~\cite{10.1063/1.365356}
    \textsuperscript{i}~\cite{tamargo2002ii}
    \textsuperscript{j}~\cite{Poon1995RelativisticBS}
    \textsuperscript{k}~\cite{10.1063/1.2899087}
    \textsuperscript{l}~\cite{SWEYLLAM2010681}
    \textsuperscript{m}~\cite{10.1063/1.1368156}
    }
    \caption{For Group III-V and II-VI semiconductors, calculated and collected experimental values of spin-orbit splitting and band gap (in eV) are listed, along with the band gap type (Direct or Indirect). Our calculated spin--orbit splitting of 0.65 eV for AlSb  is slightly lower than previously reported values of 0.74 eV~\cite{Al-Douri2013}.}
    \label{table_so2}
\end{table}
The band structure profiles, particularly the highest valence and lowest conduction bands, exhibit similar shapes. However, the length of the one-dimensional paths of high-symmetry points and line segments in reciprocal space (k--path) is slightly different, which means the curvature of the electronic bands is different. For instance, according to Table~\ref{tab:lattice_parameters}, the computed largest lattice constant is 6.5414~\AA~for  InSb, while the shortest lattice constant is 5.3615~\AA~for ZnS. The lattice constant of ZnS is 18\%  smaller than the lattice constant of InSb, while in reciprocal space, the length of the k--path of the ZnS is 22\% larger than the length of the k--path of the InSb.  As the size of the cell in real space increases, the ﬁrst Brillouin zone in reciprocal space shrinks and vice versa~\cite {Mayo_2020}.  The effective mass is inversely correlated with the curvature of the electronic dispersion in reciprocal space~\cite{PhysRevB.99.085207}.  However, in this study, we emphasize that there is no apparent direct relation between the effective mass of the electron and the shift current.

Table~\ref{table_so2} summarizes the nature of the band gaps and their values in the zincblende semiconductors studied in this work.  For semiconductors AlAs, AIP, AlSb, GaP, and ZnS, the band gaps are characterized as indirect, with the valence band maximum at $\Gamma$ point  and the conduction band minimum at X point. In contrast, other compounds exhibit a band gap direct at  $\Gamma$ point.  According to our theoretical predictions, as shown in Table~\ref{tab:lattice_parameters}, the computed band gap values range from 0.05 eV to 4.30 eV.  The computed band gaps of semiconductors  InP, CdS,  CdSe, CdTe, and ZnS are overestimated. In contrast,  the computed band gaps of semiconductors  AlAs, AIP, AlSb, GaAs, GaP, GaSb, InAs, InSb, and ZnTe are underestimated comparatively to the experimental data.  A closer analysis of Table~\ref{table_so2} shows that  the computed band gap for GaAs is 1.2 eV,  which is 21\%  underestimated compared to the experimental data of band gap with the value of 1.519 eV for GaAs at  0 K~\cite{10.1063/1.1368156}.  Note that we are using the lattice constant (PBEsol) of 5.6836~\AA~which is overestimated by {0.53\%} compared to the experimental value. Under these conditions, the calculated band gap yields 1.2 eV.  Interestingly, we also consider the lattice constant of GaAs to  5.6241~\AA,~which is underestimated by {0.51\%}, relative to the experimental data.  With this lattice constant, the computed band gap for GaAs yields 1.46 eV,   which is closer to the experimental value of 1.519 eV.

In the case of the InSb semiconductor, we utilized the lattice constant obtained from the PBEsol method, which is 6.5414~\AA.~This value is overestimated by {4\%} compared to the experimental data. When using this lattice constant, the calculated band gap is 0.05 eV, which is roughly 78\% underestimated relative to the experimental band gap, as shown in Table~\ref{table_so2}.  However, when we use the lattice constant of 6.4711~\AA,~which is slightly underestimated by {0.13\%},  the band gap calculation yields 0.26 eV, which is closer to experimental data.  Notably, if we calculate the band gap of a semiconductor using an overestimated lattice constant, this would yield a significant deviation in the band gap value. In contrast, if we use an underestimated lattice constant, it yields a better correction of the band gap.  Based on our calculations, we found that the value of the band gap is somewhat sensitive to slight variations in the lattice parameter when employing the mBJ approach to compute electronic properties.  We underline that our predicted band gaps do not take into account the zero-point renormalization of the band gap~\cite{PhysRevB.106.085137}.  Overall, our energy band gaps agree reasonably well with the available experimental and theoretical results, and the mBJ scheme provides an efficient framework for band gap calculation in zincblende semiconductors.

\subsection{Shift current response in zincblende GaAs semiconductor}
\begin{figure}[ht!]
  \includegraphics[scale=0.85]{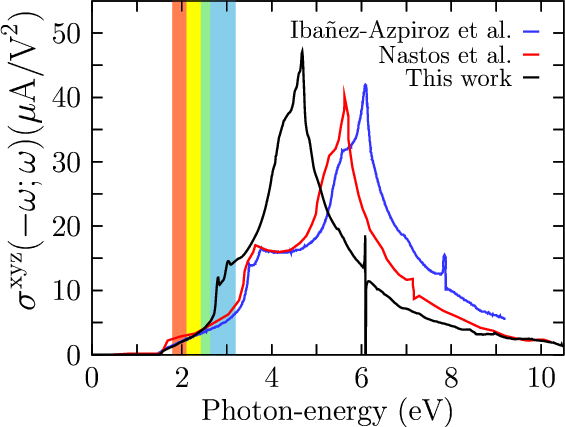}
  \caption{(color online) We show the comparison of our calculated spectrum of the shift current tensor [$\sigma^{xyz}(0;\omega,-\omega)$] for zincblende GaAs with results obtained by  Iba\~nez-Azpiroz et al.~\cite{PhysRevB.97.245143} and Nastos et al.~\cite{Nastos2006}. All spectra include scissors corrections to match the 1.519 eV band gap of GaAs. The rainbow bar shows the visible light spectrum, ranging from about 1.8 to 3.1 eV~\cite{OZCAN2007291}.}
  \label{shift0}
\end{figure}
At the beginning of the year 2000, Sipe et al. presented a calculation of the spectrum of the shift current tensor, $\sigma^{xyz}(0;\omega,-\omega)$, for GaAs~\cite{Sipe2000}. Their approach was based on the principles of nonlinear optics. Around the same time, Petr et al. also reported findings on the shift current in GaAs semiconductors using the framework of Green’s functions \cite{PetrKral_2000}.

The shift current  tensor $\sigma^{abc}(0;\omega,-\omega)$   is a third-order tensor that consists of 27 components. The zincblende symmetry of GaAs requires that the third rank tensors must be symmetric under the exchange of any pair of indices~\cite{Nastos2006}, which means that  $\sigma^{abc}(0;\omega,-\omega) = \sigma^{acb}(0;\omega,-\omega)$.  The shift current is prohibited only in crystal class 432~\cite{Sipe2000}  and is present for GaAs and other semiconductors that lack a center of inversion symmetry.

\begin{figure*}[ht!]
  \includegraphics[scale=0.82]{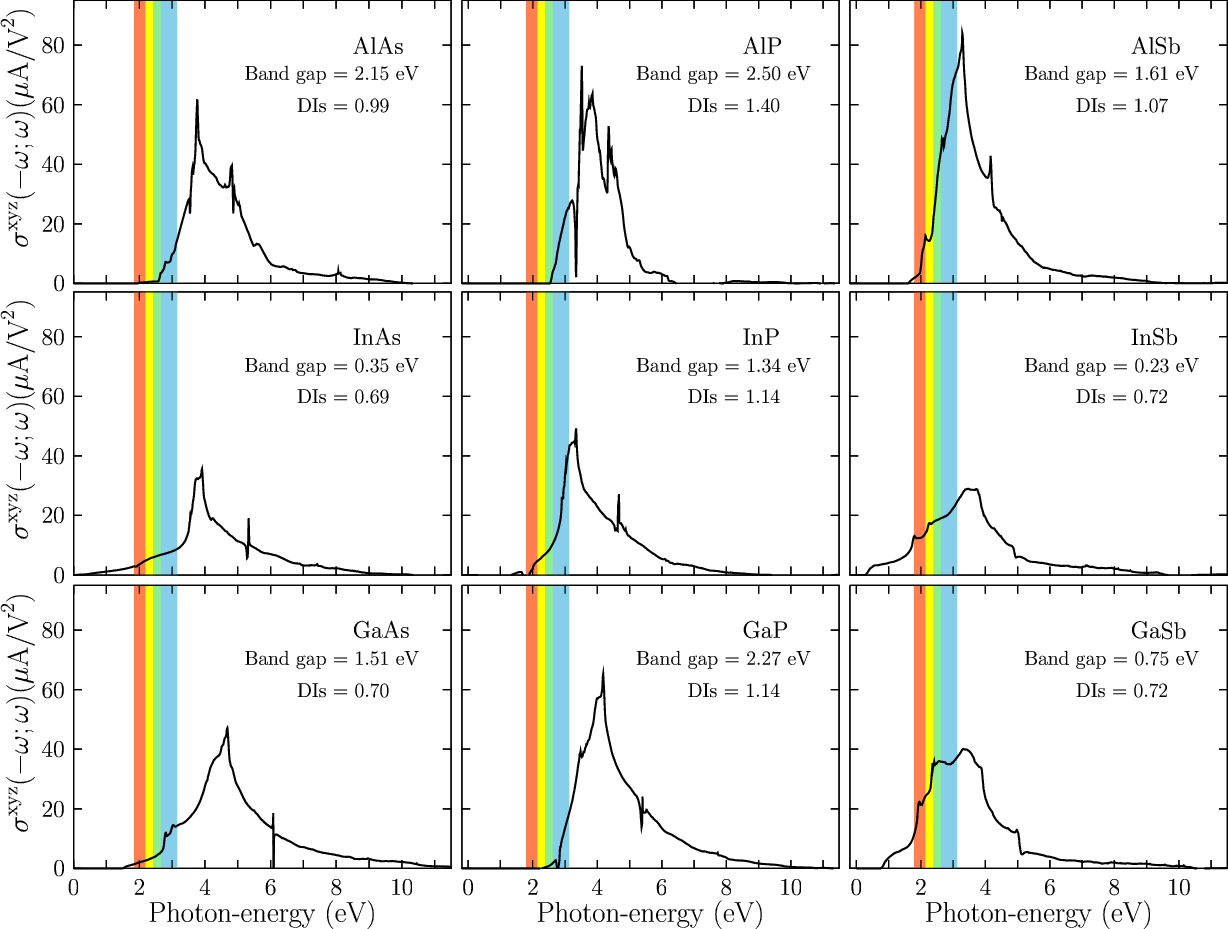}
  \caption{(Color online)  The spectra of the shift current tensor component xyz [$\sigma^{xyz}(0;\omega,-\omega)$]  for zincblende semiconductors belonging to group III--V. Among all Group II--VI and III semiconductors, AlSb semiconductor exhibits the highest peak of shift current when exposed to visible light illumination. Visible light, depicted as a rainbow zone, spans a range of energies between 1.8 and 3.1 eV~\cite{OZCAN2007291}. DIs refer to the computed delocalization indices.}
  \label{shift1}
\end{figure*}

Figure~\ref{shift0} shows our computed spectrum of the shift current tensor component xyz [$\sigma^{xyz}(0;\omega,-\omega)$] or shift current spectrum represented by the solid black line.  For easy comparison,  in Figure~\ref{shift0}, we have also included  the calculation of shift current spectrum reported by Iba\~nez-Azpiroz et al.~\cite{PhysRevB.97.245143} depicted by a solid blue line and the calculation of shift current spectrum reported by  Nastos et al.~\cite{Nastos2006} depicted by the solid red line. In the spectra shown in Figure~\ref{shift0}, different values of scissors correction were applied, considering a band gap of 1.519 eV for GaAs~\cite{10.1063/1.1368156}. It is important to note that for a linear optical response, the scissors correction rigidly shifts the spectrum upward along the energy axis without altering its shape~\cite {PhysRevB.80.155205, PhysRevB.72.045223},  and the shift current spectrum also undergoes these rigid energy shifts~\cite{PhysRevB.80.245204,PhysRevB.97.245143}.  Iba\~nez-Azpiroz et  al.~\cite{PhysRevB.97.245143} utilized a Wannier-Interpolation scheme for calculating the shift current spectrum. In contrast, Nastos et al.~\cite{Nastos2006} presented a shift current spectrum for GaAs in the context of nonlinear optics, employing a full--band structure calculation. We adopt the methodology of Nastos et al and T. Rangel et al. to compute the shift current spectrum. Our calculated shift current spectrum, depicted by the solid black line, and displayed in Figure~\ref{shift0}   are in reasonable agreement with those spectra reported by  Iba\~nez-Azpiroz et al.~\cite{PhysRevB.97.245143} and Nastos et al.~\cite{Nastos2006}.

Our computed shift current spectrum exhibits the largest peak, with a value of 47 $\mu \mathrm{A/V}^2$, located at a photon energy of 4.5 eV. In contrast, the shift current spectra reported by Iba\~nez-Azpiroz et al. and Nastos et al. present the largest peaks with values of 43 and 41 $\mu \mathrm{A/V}^2$, respectively, located at 6.0 and 5.5 eV.  The value of the largest peak in our computed shift current spectrum is slightly larger, approximately 10\%, compared to the largest peak of shift current previously reported by Ibañez-Azpiroz et al. and Nastos et al. Another previous study  reported a computed shift current spectrum that exhibited a maximum peak with a value of 20 $\mu \mathrm{A/V}^2$ for the zincblende GaAs semiconductor; however, the spin-orbit effect was not considered by the authors~\cite{D4CP02478H}.

We point out that there is minimal overlap between the shift current spectra of GaAs and the visible light spectrum, as shown in Figure~\ref{shift0}. Consequently, the shift current generation in zincblende GaAs semiconductors is limited to below 15 $\mu \mathrm{A/V} $  in the visible region.

\subsection{Shift current response in zincblende  III-V and II-V semiconductors}

For ease of comparison,  Figure~\ref{shift1} displays the spectra of the shift current tensor component xyz [$\sigma^{xyz}(0;\omega,-\omega)$] for zincblende semiconductors belonging to group III-V. The spectra of the shift current tensor component xyz  [$\sigma^{xyz}(0;\omega,-\omega)$ ] for the zinc chalcogenides (ZnS, ZnSe, and ZnTe) and cadmium chalcogenides (CdS, CdSe, and CdTe) are shown in Appendix~\ref{appendix:b}, Figure~\ref{shift2}. It is important to note that both Figures show only the non--zero xyz component. Furthermore,  each panel in Figure~\ref{shift1} and  Figure~\ref{shift2} is 
identified with its specific chemical stoichiometry. At first glance,  the spectra of the shift current tensor [$\sigma^{xyz}(0;\omega,-\omega)$],  shown in Figure~\ref{shift1}, exhibit quite analogous shapes with a strong peak over a broad span of energy,  except for some minor spectral deviations.
Our findings reveal  that the maximum peak  of the shift current spectrum in aluminum, indium, and
gallium pnictides (AlSb, AlP,  AlAs,  InSb, InP, InAs,  GaSb, GaP, and GaAs) are 83, 70, 61, 29, 49, 35, 40, 60,
and 47  $\mu \mathrm{A/V}^2$, respectively located at the photon energies of 3.1, 3.5, 3.8, 3.4,4.3, 4.4, 3.3,4.5 and 4.6 eV respectively.
Importantly, the zincblende AlSb semiconductor exhibits the highest shift current peak 
with the value of 83  $\mu \mathrm{A/V}^2$  in response to visible light illumination
 surpassing the zinc chalcogenides, cadmium chalcogenides, and aluminum, indium, and gallium pnictides.
 In order for a material to exhibit a significant shift current response from sunlight and thus be effective
 for solar energy harvesting material,  it needs to possess a band gap within the visible spectrum or the near-infrared~\cite{10.1063/1.4901433}. The zincblende AlSb semiconductor possesses an observed band gap of  1.61 eV,  which is slightly below the lowest value of the visible light spectrum.

 The shift current spectrum of AlSb is shown in the upper right panel of  Figure~\ref{shift1}. At the onset of the signal, just slightly above the indirect band gap energy, the spectrum rises almost vertically, reaching a maximum peak of 83 $\mu \mathrm{A/V}^2$, which occurs at a photon energy of 3.1 eV.   At higher energies, the shift current spectrum intensity rapidly decreases exponentially and almost vanishes at photon energies of  7 eV. It is important to note that the shift current spectrum significantly overlaps with the visible light spectrum; consequently, the zincblende AlSb semiconductor can be excited by visible light. This makes the zincblende AlSb semiconductor a promising candidate for photovoltaic applications. In contrast, the zincblende CdSe semiconductor exhibits the lowest induced shift current spectrum as displayed in Appendix~\ref{appendix:b}, Figure~\ref{shift2}.  The most prominent peak in the shift current spectrum of CdSe has a value of 14 $\mu \mathrm{A/V}^2$, located at the photon energy of 7 eV. Furthermore, the shift current spectrum of CdSe falls outside the visible light. 
 
 Our computed  shift current spectrum  in the zincblende AlSb is an order of magnitude to recent experimental observations of shift current in ferroelectric semiconductor  SbSI~\cite{doi:10.1073/pnas.1802427116} and also the shift current computed in the ternary compound ZnSnP, which is a photo absorber material with solar cell application~\cite{PhysRevMaterials.4.064602}.
This shift current in the zincblende AlSb (83 $\mu \mathrm{A/V}^2$) is higher than the shift current in PbTiO$_3$ and
BaTiO$_3$ where the shift currents generated by those materials are below (10 $\mu \mathrm{A/V}^2$)~\cite{PhysRevB.100.245206, PhysRevLett.109.116601}.  Moreover, the shift current computed in AlSb exceeds that of the shift currents calculated in wurtzite and zincblende semiconductors reported in recent studies~\cite{D4CP02478H}. A previous study reported that single-layer Ge and Sn monochalcogenides exhibit shift current with a value of 100 $\mu \mathrm{A/V}^2$~\cite{PhysRevLett.119.067402} which is roughly {20\%} higher  than our computed  shift current value of the 83 $\mu \mathrm{A/V}^2$ in zincblende AlSb semiconductor.   Recently published work reports the theoretical shift current spectrum in the SnTe monolayer exhibits a high peak of value 284 $\mu \mathrm{A/V}^2$ located at a photon energy of 1.21 eV~\cite{Jin2024}.
\begin{figure}[htb!]
  \includegraphics[scale=0.85]{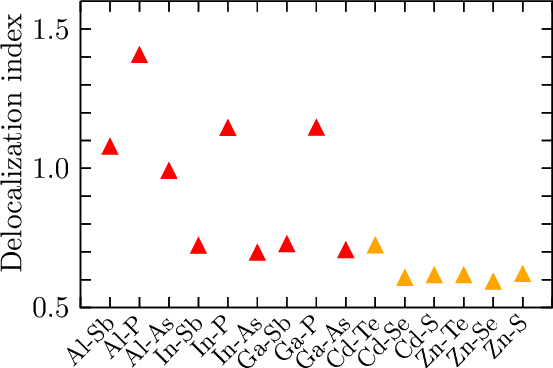}
  \caption{(Color online) We present the delocalization indices calculated between the first neighbors for the aluminum, indium, and gallium pnictides (red, filled triangles), as well as for zinc chalcogenides and cadmium chalcogenides (orange, filled triangles). The x--axis represents chemical bonding, and the y--axis indicates delocalization indices.}
  \label{loca}
\end{figure}
The value of 284 $\mu \mathrm{A/V}^2$ is 3.4 times larger than the shift current of 83 $\mu \mathrm{A/V}^2$ exhibited in the zincblende AlSb semiconductor. However, it is important to note that the shift current spectrum of the SnTe monolayer falls outside the visible spectrum.

All the shift current spectra shown in Figure~\ref{shift1} exhibit a similar peak structure, although there are variations in intensity.  Previous studies determined that the geometries of the energy band structures determine the shift current~\cite{PhysRevResearch.6.013123,PhysRevLett.109.116601}. At first glance,  the band structures displayed in Figure~\ref{bands1} are identical in shape. As a consequence, the shift current spectra shown in Figure~\ref{shift1} are similar.

The highest shift current in AlSb semiconductor can be attributed to several factors, including a small band gap and the SOC effect.    There are several reasons to take into account the SOC  in the computation of band structure and shift current response. Prior studies highlight that the inclusion of the SOC effect causes drastic changes in the shift current spectra~\cite{Qian2023,D4CP02478H}. Moreover, calculations that exclude SOC may result in spin splitting being omitted, leading to an inaccurate description of the electronic band structure and band gap, as well as significant changes in electronic transport properties~\cite{GU2023112090}.

The data presented in Table~\ref{table_so2} indicates that the zincblende AlSb semiconductor,
which generates the largest shift current,
has a computed spin--orbit splitting value of 0.65 eV. In contrast,  the zincblende CdSe semiconductor, which
generate the lowest shift current, has a spin-orbit splitting of 0.40 eV. Additionally,  a closer examination of
Table~\ref{table_so2} reveals that the zincblende AlP semiconductor, which generates the second large 
shift current, has a computed spin-orbit splitting of only 0.05 eV.

These observations suggest that there is no clear correlation between the shift current and the values of spin--orbit splitting.   To assess the impact of SOC effects on the shift current, both with and without SOC, we refer to Figure~\ref{shift-SOC} in Appendix~\ref{appendix:c}. This Figure compares the spectra of the shift current tensor both with and without SOC for several semiconductors, illustrating that the spectra remain nearly unchanged when SOC is present or absent. Therefore, the absence of  SOC has a minimal effect on the shift current spectrum for those zincblende semiconductors considered in this study.

Previous work found that the effect of SOC on the calculated dynamics of the shift current induced on the WS$_2$ monolayer has a minimal effect~\cite{PhysRevResearch.6.013123}. In contrast, prior studies have shown that including the SOC effect can nearly double the maximum value of the shift current~\cite{D4CP02478H}. Other previous investigations on shift current in two--dimensional ferroelectrics show that the inclusion of the SOC effect could significantly impact the interband Berry connections and lead to substantial changes in the shift current spectra~\cite{Qian2023}.  In conclusion,  the SOC effect should be considered in all calculations related to the shift current.

Referent to band gap, previous studies indicate that a reduced band gap significantly enhances the shift currents in 2D ferroelectrics~\cite{Qian2023}. The largest peaks of the shift current spectra displayed in Figure~\ref{shift1} belong to the zincblende  GaP, AlP, AlAs, and AlSb semiconductors with computed indirect electronic bandgaps 2.27, 2.09,  1.87, and 1.35 eV respectively.  A close analysis of Table~\ref{table_so2}, we see that some semiconductors have larger band gaps, namely 4.30 eV for ZnS, 3.03 eV for CdS, and 2.93 eV for ZnTe. In contrast, some semiconductors have smaller band gaps: 0.05 eV for InSb, 1.20 eV for GaAs, and 1.30 eV for InAs. Our results indicate that semiconductors with the smallest band gaps do not generate the largest shift current. 

The shift current tensor can be expressed as the product of the shift vector and the optical transition intensity, as shown in Equation~\ref{shift_current_response2}. We specifically focus on the delocalization of electronic states since optical transitions involve both an initial and a final band. Figure~\ref{loca} shows the delocalization Indices (DIs) of zincblende III--V (AlP, AlAs, AlSb, GaP, GaAs, GaP, InP, InAs, and InSb) and II--VI (ZnS, ZnSe, ZnTe, CdS, CdSe, and CdTe) semiconductors. Most important of all, the zincblende AlSb, AlP, AlAs, and GaP semiconductors show larger values of DIs with  1.07, 1.14,  0.99, and 1.14, respectively,  while the cadmium chalcogenides and zinc chalcogenides shows the lower values of  DIs with 0.72,  0.60,  0.61, 0.61, 0.59,  0.62 respectively. Remarkably, our calculations reveal a quantitative correlation between the DIs and the intensity of the generated shift current in the zinc blende semiconductors studied here. Therefore, it elucidates the fact that the shift current is more sensitive to the nature of bonding, covalency effects~\cite{PhysRevMaterials.8.025001}, and delocalization of electronic states than to the value of the band gap or the inclusion of the SOC effect. In bonds formed by two atoms sharing an electron pair, the DIs between the two atoms take values  close to 1.0 for an almost complete shared electron and close to zero for an ionic bond. The results in Figure~\ref{loca} indicate that the zinc chalcogenides (ZnS, ZnSe, and ZnTe) and cadmium chalcogenides (CdS, CdSe, and CdTe) are more ionic materials. In contrast, zinc blende AlSb,  AlP, AlAs, and GaP are more covalent materials.   The DIs are larger when more aluminum elements are present in the solid's composition. As a consequence,  the materials with aluminum present in their composition tend to generate a larger shift current in the zincblende semiconductors studied in this work.  On the other hand, we found that phosphorus also increases the response, but its effect is less significant compared to aluminum.  Importantly, according to the spectra shown in Figure~\ref{shift1},  materials containing  SbX (X=Al, In,  Ga) generate shift current spectra that overlap entirely with the visible light spectrum.  We suggest that materials composed of Sb elements will generate a shift current spectrum within the visible light spectrum. The relationship between the magnitude of the shift current and chemical species is unclear, necessitating further investigation. Nevertheless, our results are in the same direction as previous studies that delocalized electronic states enhance shift current eﬀects~\cite{PhysRevB.103.245415, Tan2016, Tan_2019,PhysRevB.100.085102}. 

\subsection{Band--by--band decomposition of shift current spectrum of zincblende AlSb semiconductor}
We now turn our attention to the breakdown of the spectrum of shift current tensor $\sigma^{abc}\left(0;\omega,-\omega\right)$ of zincblende AlSb semiconductor. The calculation of shift current spectrum is carried out under the sum-over-states  formalism using the Equation~\ref{sigma1} that depends on the position matrix elements within the r · E gauge~\cite{PhysRevB.80.155205}  and the energy band structure differences ~\cite{PhysRevB.70.235110}.  The shift current spectrum can be broken down and examined by summing only selected valence and conduction bands. This method enables us to investigate the distinct contributions of different bands to the shift current spectrum in zincblende semiconductors.

\begin{figure}[htb!]
  \includegraphics[scale=0.80]{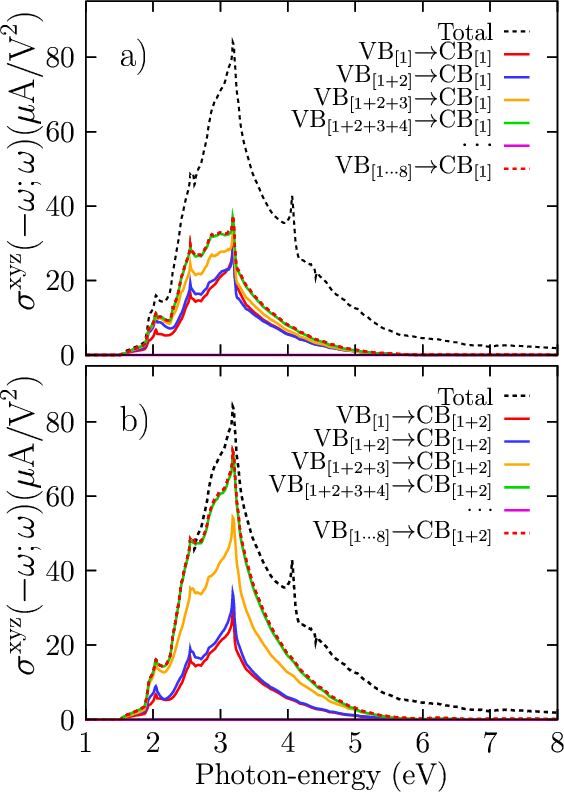}
  \caption{(Color online) Breakdown of the shift current spectrum in its electronic band's contributions in the zincblende AlSb semiconductor. Transitions involve the heavy--hole, light--hole, and split--off valence bands.  The split--off valence bands make zero contribution to the shift current spectrum.  The optical transitions are identiﬁed in Figure~\ref{trans}. 
}
  \label{shift5}
\end{figure}

Figure~\ref{shift5} shows the band--by--band contributions to the shift current response for the separate transitions,  heavy-hole (VB$_{[1+2]}$),   the combined heavy--hole plus light--hole  (VB$_{[1+2+3+4]}$), and split off  (VB$_{[4+6]}$) plus heavy--hole, light--hole and the deeper energy valence bands. The upper panel a) in Figure~\ref{shift5} displays the excitations into the lowest conduction band  (CB$_{[1]}$), resulting in transitions from the heavy--hole,   light--hole,  and split--off bands. Panel a) shows the shift current spectrum that involves  the transitions  from
heavy--hole to the first  conduction band, and it is depicted by
blue solid line  (VB$_{[1+2]}$$\rightarrow$CB$_{[1]}$ ),
the largest peak in the shift current spectrum has a value of approximately 25 $\mu \mathrm{A/V}^2$.  It is noteworthy that 
transitions from the first valence band to the first conduction band (VB$_{[1]}\rightarrow$CB$_{[1]}$ ), (red line) 
make a larger contribution to the shift current spectrum that heavy-hole transitions to the first conduction bands  (VB$_{[1+2]}\rightarrow$CB$_{[1]}$ ).  Panel a)  illustrates the shift current spectrum for transitions originating from the heavy hole plus a valence band that belongs to the light hole to the first conduction band (VB$_{[1+2+3]}\rightarrow$CB$_{[1]}$ ), depicted by the yellow curve.  The inclusion of a valence band that belongs to the light--hole increases the shift current spectrum maximum peak to a value of approximately 30 $\mu \mathrm{A/V}^2$.  Panel a) shows the shift current spectrum  for transitions that promote electrons from the heavy--hole  plus  light--hole  bands to the first conduction band (VB$_{[1+2+3+4]}$$\rightarrow$CB$_{[1]}$)
 is depicted by the green line. The shift current spectrum for transitions originating from the heavy--hole,  light--hole,  split--oﬀ bands, and any other valence band to the first conduction band (VB$_{[1...8]}$$\rightarrow$CB$_{[1]}$) is depicted by the dotted red curve.
\begin{figure}[htb!]
  \includegraphics[scale=1.0]{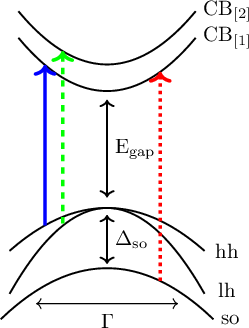}
  \caption{(Color online) Schematic representation of the  band structure of zincblende AlSb semiconductor near the $\Gamma$  point. Optical transitions are identiﬁed by arrows as follows: solid blue, dashed green,  and dotted red curves correspond to the excitation of electrons from the heavy (hh), light (lh), and split-oﬀ (so) valence bands, respectively, to the first and second conduction bands.  The optical transitions, heavy hole plus light hole (VB$_{[1+2+3+4]}\rightarrow$CB$_{[1+2]}$)   correspond to the addition of solid blue and dashed green curves.  Transitions from split-off valence bands and lower-energy valence bands do not contribute to the overall shift current spectrum. (E$_{\mathrm{gap}}$=1.35 eV,  and  $\Delta_{\mathrm{so}}$=0.65 eV for zincblende AlSb semicondutor).}   
  \label{trans}
\end{figure}
It is important to note that this shift current spectrum overlaps with the shift current spectrum generated by the transitions VB$_{[1+2+3+4]}\rightarrow$CB$_{[1]}$).  This means that the transitions from the split-off band and any other lower energy valence band make no contributions to the overall shift current spectrum.  Panel a) shows the total shift current spectrum depicted as a dotted black line. In summary, the consideration of the transitions from all valence bands to the first conduction band contributes  less than 50\% to the overall shift current spectrum, as displayed in Figure~\ref{shift5}(a). Figure~\ref{shift5}(b)  shows the excitations into the two lower-energy conduction bands  (CB$_{[1+2]}$).  from the heavy-hole,   light-hole,  and split-off bands. The shift current spectrum that includes electronic transitions from heavy-hole to the first and second conduction bands is represented by the blue solid line  (VB$_{[1+2]}\rightarrow$ CB$_{[1+2]}$ ). The peak of this spectrum is approximately 35 $\mu \mathrm{A/V}^2$, located at the energy of 3.1 eV.   Further, the shift spectrum resulting from transitions that include both heavy-hole plus and light-hole to the first and second conduction bands  (VB$_{[1+2+3+4]}\rightarrow$ CB$_{[1+2]} $) is shown by the green line.   It is important to note that those transitions from heavy-hole and light-hole states make the largest contribution to the overall shift current spectrum.   In contrast,  transitions from the split-off  and lower energy  valence bands to the first and second conduction bands (VB$_{[1...8]}\rightarrow$CB$_{[1+2]} $)  do not affect the overall shift current spectrum. The transitions are identiﬁed in Figure~\ref{trans}.  To complete the total shift current specrum we need to take into acount
higher energy conduction bands.  It is clear that the SO effect  not strong influence  ether the magnitude not the shape of the
shift current specrum, as we can see in Figure~\ref{shift-SOC} due to the   the dominant contribution arises from the first four valence bands (heavy--hole and light--hole), accounting for the majority of the response, while the remaining valence bands contribute almost zero.   Full band structure calculations  gives access to the full  k-path valleys at the  $\Gamma$ ,  X,  K , and  L symmetry points.
The contribution from the $\Gamma$ --X region of the Brillouin zone yields the largest peak in
shift current spectra  displayed in Figure~\ref{shift5}(b) while  the contribution to the onset of the shift currrent spectrum  comes from  $\Gamma$  point.
\begin{figure}[htb!]
  \includegraphics[scale=0.54]{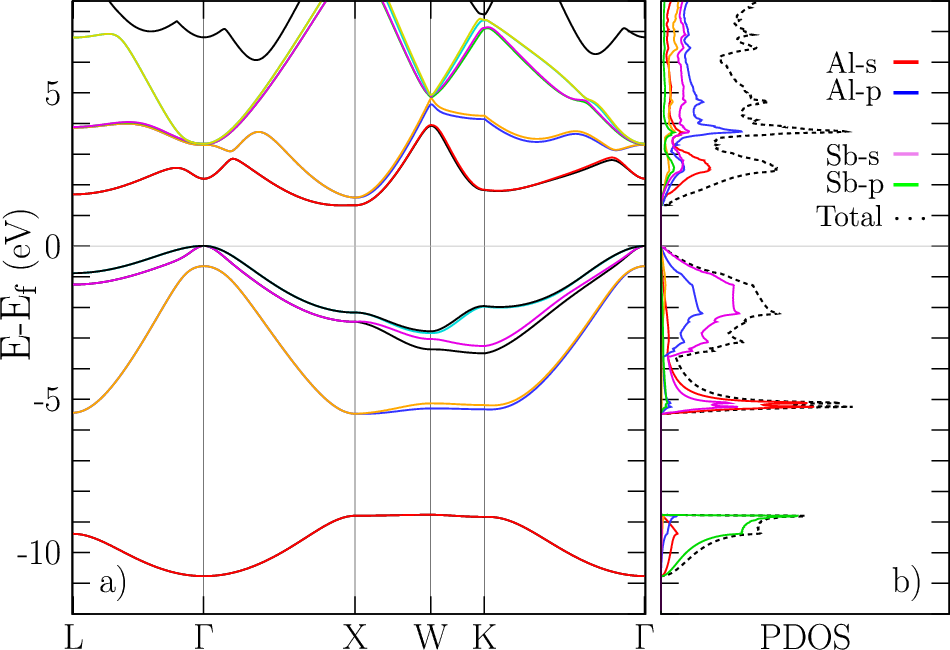}
  \caption{(Color online) The left panel displays the calculated electronic band structure, and the right panel shows the partial density of states of the zincblende AlSb semiconductor. The Fermi level is shifted to zero. The heavy--hole and light--hole bands are formed predominantly by different contributions of the  Al--p and  Sb--s states.}
  \label{fpdos}
\end{figure}
For a deeper insight, we compute the partial density of states (PDOS) of zincblende AlSb semiconductor. 
Figure~\ref{fpdos}, left panel, presents the band structure  and, right panel,  its corresponding PDOS.
Due to the shared structure and similar band structures, we present only the density of states for AlSb within the energy interval of 12 to 8 eV.
The density of states of the other semiconductors is expected to exhibit minimal variations since their band structures are comparable.
The uppermost valence bands, heavy and light--hole,  situated near the Fermi level,  in  energy range  0  to -3 eV, primarily consist of the
hybridization of Sb--s and Al--p states.
The first two conduction bands are located around  in the energy range
1.35 to 3.5 eV and they are  predominantly composed of mixture  of Al--p and Sb--s states and a small aomunt  of Sb--p states.
The onset and the largest peak of the total shift current spectrum, displayed in Figure~\ref{fpdos},   is due to the
direct transitions involving mainly Sb--s states  and  Al--p states of the valence bands.   
\section{Effect of strain  on the shift current  response}
A strategy to analyze changes in the shift current response generated by the zincblende AlSb semiconductor is to cause deformation through hydrostatic pressure. In this study, we also investigated the effect of hydrostatic strain on the shift current response. Specifically, we apply positive hydrostatic strain (tensile)  and negative hydrostatic strain (compressive) to zincblende AlSb semiconductors, which exhibit the highest shift current generation under uniform illumination.
  \begin{figure}[H]
  \includegraphics[scale=0.85]{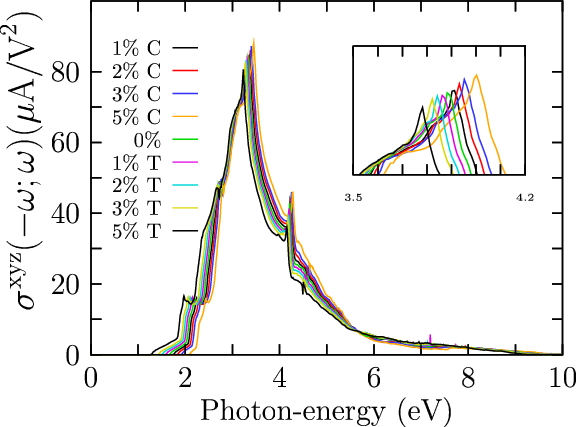}
    \caption{(Color online)The effect of hydrostatic strain on the shift current spectrum in the zincblende AlSb semiconductor.}
  \label{fstrain}
\end{figure}
Figure~\ref{fstrain} shows the shift current spectra for both compressive and tensile strain in the zincblende AlSb semiconductor. We found that a tensile strain of {5\%} slightly enhanced the largest peak to 86  $\mu \mathrm{A/V}^2$  in the shift current spectrum; in contrast, a compressive strain slightly reduced the largest peak to  80  $\mu \mathrm{A/V}^2)$ in the shift current spectrum. According to Figure~\ref{fstrain}, one notices that the compressive and tensile strains do not cause drastic changes in the shape or the magnitude of the shift current spectra. The onset of the signals is shifted in the energy scale according to the value of the electronic band gap.  According to the data in Table~\ref{table_pressure},  a tensile strain (+) increases the band gap and reduces the SOC; in contrast, a compressive strain reduces the indirect band gap and increases the SOC. However, the direct band gap at the $\Gamma$  point increases. The indirect ($\Gamma$--X) bandgap decreases with increasing pressure. This behavior is due to the compression of atomic distances, which can enhance the overlap of atomic orbitals , thereby increasing the charge transfer~\cite{ALGARNI2018215}.  The shift current spectra, displayed in Figure~\ref{fstrain}, of the zincblende AlSb semiconductor are not sensitive to either the change in the band gap value or pressure.    We attribute this insensitivity to the fact that the shape of the band structure, even for different strain values,  remains nearly identical to that calculated at the equilibrium volume.
\begin{table}[ht!]
    \centering
    \renewcommand{\arraystretch}{1.2}
    \setlength{\tabcolsep}{4pt}
    \begin{tabular}{lllc}
         \toprule
        \textbf{ Pressure.} & \hspace{1mm} \textbf{SO effect (eV)} & \hspace{1mm} \textbf{Band gap (eV)}  \\
         \midrule
        -5\%   & \hspace{3mm}  0.6638 & \hspace{3mm} 1.2763 \\
        -2\%   & \hspace{3mm}  0.6543 & \hspace{3mm} 1.3236  \\
        -1\%  & \hspace{3mm}   0.6513 & \hspace{3mm} 1.3377  \\
        \Large{e$_0$}  & \hspace{3mm} 0.6483  & \hspace{3mm}  1.3509 \\
        +1\%   & \hspace{3mm}  0.6455  & \hspace{3mm}  1.3635\\
        +2\% & \hspace{3mm}  0.6427 & \hspace{3mm}  1.3753 \\
        +5\%  & \hspace{3mm} 0.6347 & \hspace{3mm}  1.4070 \\
        \bottomrule
    \end{tabular}
    \vspace{1mm}
    \caption{The electronic band gap and  spin-orbit splitting effect  values computed  as a function of hydrostatic pressure for zincblende AlSb semiconductor. The band gap narrowing and SOC increase as increasing pressure, In contrast, when tensile strain is aplied, the band gap increase and SOC values decrese.}
    \label{table_pressure}
\end{table}
We noted that the behavior of the shift current under pressure may differ in other materials, which are distinct from those examined in this study.  Previous work applied hydrostatic pressure to topological insulator BiSb and found significantly enhanced shift current~\cite{doi:10.1021/acs.jpcc.4c03673}.   Furthermore, another previous theoretical investigation noted that the magnitude of the shift current increases in two--dimensional ferroelectric monolayers~\cite{Qian2023}.
\section{Conclusions}
In this study, we investigated fifteen zincblende semiconductors to determine which one generates the highest shift current under uniform illumination and investigated the underlying mechanisms that enable this effect.   We identified AlSb as the material exhibiting the highest peak in the shift current spectrum, achieving a value of 83 $\mu \mathrm{A/V}^2$.  Notably,  the shift current spectrum falls entirely within the visible light spectrum. At the same time, the AlP is the second semiconductor with a higher shift current response. In contrast, CdSe and CdS materials demonstrated lower shift current responses. 

Notably, materials containing aluminum (Al)  show a significantly high shift current response. However,  the antimony (Sb) atom plays an important role in enhancing the shift current and the incusion of Sb atom makes to the material tend to generate shift current spectra within  the visible light spectrum.  Importantly, we found a clear correlation between the delocalization indices and the shift current. In contrast, the spin--orbit splitting values and the band gap values do not show a clear trend with shift current.

In addition, under the sum over--states scheme, we break down the shift current spectrum of the AlSb semiconductor into contributions from selected valence and conduction bands. Thus, we identified the bands responsible for the shift current spectrum, specifically those bands composed of aluminum and antimony states.

Several interesting observations emerged from our findings. Our band-resolved shift current analysis revealed that the transitions that involve only the heavy hole bands, the light hole bands, and two conduction bands account for most of the shift response signal. The Al--s electronic states mainly form the conduction bands. On the other hand, the heavy--hole bands are primarily composed of Sb--s and Al--p electronic states, as evidenced by the computed partial density of states. Therefore, the optical transitions involve Al--s, Sb--s, and Al--p states. 

Based on our band-resolved shift current analysis, the split--off bands play a minimal role in enhancing the shift current response. As a result, a larger spin-orbit splitting value does not necessarily improve the signal. For instance, in the AlSb semiconductor,  which exhibits the largest shift current response,  the computed SO splitting value is 0.65 eV. In contrast, CdSe has an SO splitting value of 0.40 eV and shows a lower generation of  shift current. Similarly, the computed SO effect for CdS is only 0.01 eV, leading to a minimal shift current. Thus, split--off bands and valence bands located energetically deeper do not significantly contribute to generating shift current.

Moreover,  we investigated the effects of hydrostatic pressure on the shift current response. We found that an increase in pressure causes the onset of the shift current spectrum to shift towards higher energies, while a decrease in pressure causes the onset and the most prominent peak of the shift current signal to shift towards lower energies. So, the increase/decrease in pressure in the AlSb semiconductor does not enhance the signal of shift current, whereas, according to literature,  in  2D materials, tensile strain increases the shift currents.

This study shows that 3D zincblende semiconductors containing aluminum (Al) and antimony (Sb) elements, such as AlAs, AlP, and AlSb, are the most promising candidates for generating a shift current. Besides, materials composed of Sb elements tend to generate shift current spectra that overlap with the visible light spectrum. For future research on materials for shift current generation within the visible light spectrum, we will explore semiconductors containing Al and Sb elements. 

\section{Acknowledgment}
E.~P.-S. (CVU 1008864) acknowledges a PhD scholarship from Mexico's  National Council of Humanities, Sciences, and Technologies, abbreviated  SECIHTI.   We gratefully acknowledge the Computational Chemistry Laboratory at the Polytechnic University of Tapachula for partially providing computational resources. 
\section{Conflicts of Interest} The authors declare that they have no conflict of interest.
\section{Funding} This research not receive funding.
\section{Abbreviations}
The following abbreviations are used in this manuscript:\\
Density Functional Theory (DFT)\\
Bulk Photovoltaic Effect (BPVE)
\appendix
\section{Band structure fo the zinc and cadmium chalcogenides}
To facilitate comparison, we have placed the band structure of zinc chalcogenides and cadmium
chalcogenides in a unique Figure. 
\label{appendix:a}
\onecolumngrid 
\begin{figure*}[htb!]
  \includegraphics[scale=0.75]{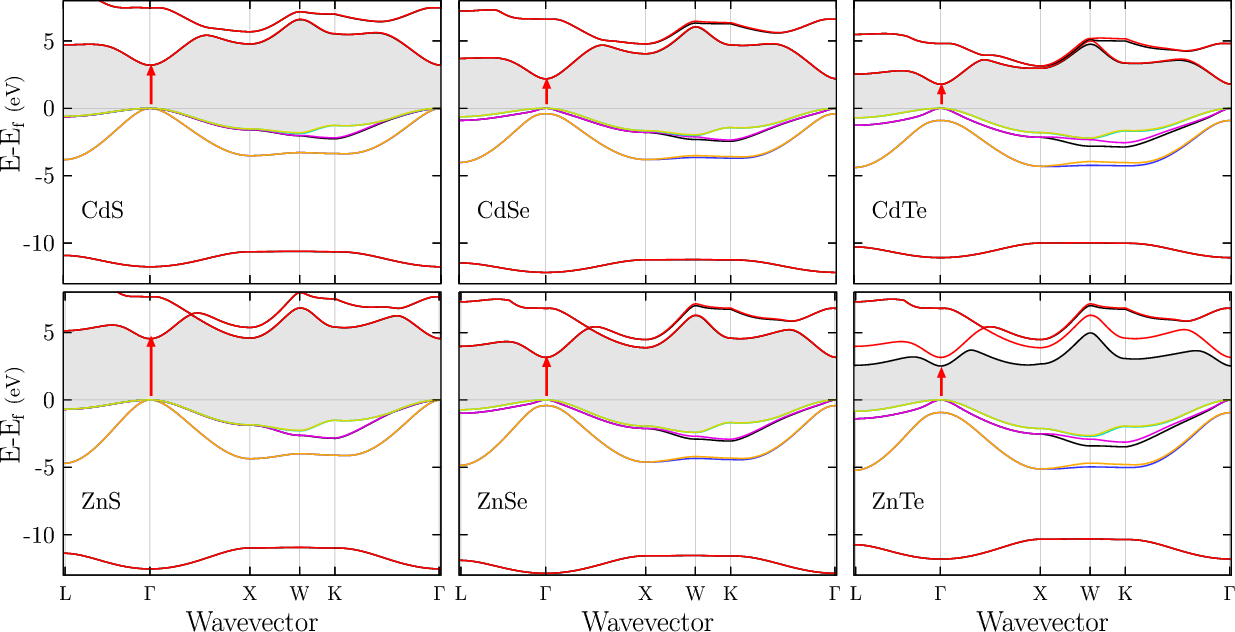}
  \caption{(Color online)  The computed band structure along with the high-symmetry directions of the Brillouin zone 
for the zinc chalcogenides (ZnS, ZnSe, and ZnTe) and cadmium chalcogenides (CdS, CdSe, and CdTe) semiconductors.  The shaded region represents the forbidden band, while the red arrows indicate the band gap. This band structure is displayed without applying the scissors correction.}
  \label{bands2} 
\end{figure*}
\section{ The shift current spectra for the ZnS, ZnSe, ZnTe CdS, CdSe, and CdTe semiconductors}
The spectra of the shift current tensor component xyz ($\sigma^{xyz}(0;\omega,-\omega)$) for zinc chalcogenides and cadmium chalcogenides are analyzed. Among all Group II-VI and III-VI semiconductors, CdSe exhibits the lowest shift current.
\label{appendix:b}
\begin{figure*}[ht!]
  \includegraphics[scale=0.8]{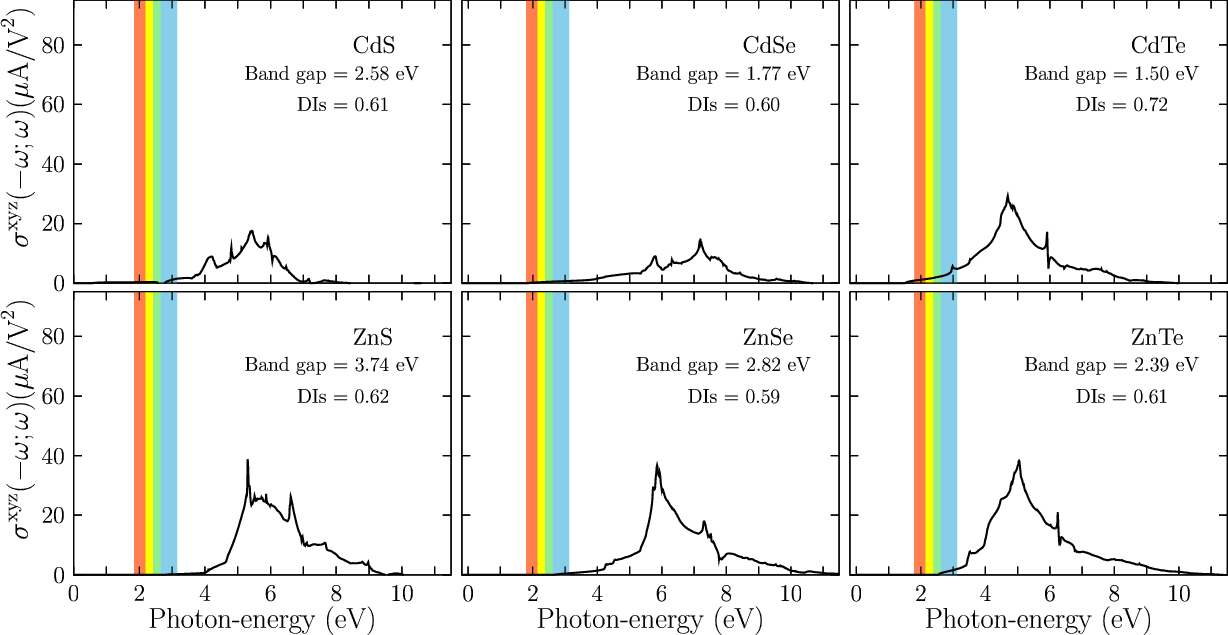}
  \caption{(Color online)  The spectra of shift current tensor, component xyz ($\sigma^{xyz}(0;\omega,-\omega)$)  for the zinc chalcogenides (ZnS, ZnSe, and ZnTe) and cadmium chalcogenides (CdS, CdSe, and CdTe) 
 The CdSe semiconductor exhibits the lowest shift current among all Group II-VI and III-VI semiconductors. Visible light appears as a rainbow and has energy levels between 1.8 and 3.1 eV\cite{OZCAN2007291}.}
  \label{shift2} 
\end{figure*}
\section{ Comparison of the shift current spectra with SOC and without SOC.}
The shift current spectra for zincblende semiconductors, including AlAs, AlP, AlSb, GaAs, ZnTe, and CdSe, are examined considering both spin-orbit coupling (SOC) and without SOC.

\label{appendix:c}
\begin{figure*}[ht!]
  \includegraphics[scale=0.8]{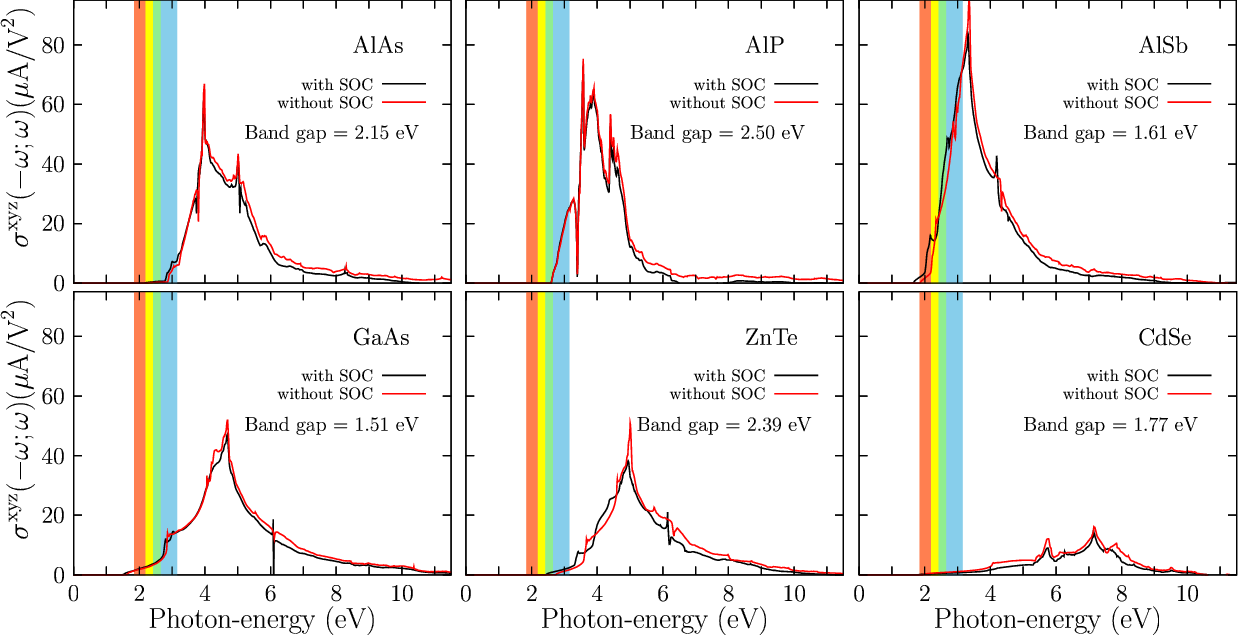}
  \caption{(Color online)  The shift current spectra for zincblende semiconductors AlAs, AlP, AlSb, GaAs, ZnTe, and CdSe, considering the SOC and without SOC.}
  \label{shift-SOC} 
\end{figure*}
\section{The $\sigma^{abc}(0;\omega,-\omega)$  tensor.}
The $\sigma^{abc}(0;\omega,-\omega)$  tensor expresd as  product of shift-vector and position matrix elements. 

\label{appendix:Rmn}
The shift current Equation~\ref{shift_current_response} after some arregements yields  Equation~\ref{shift_current_response2}. 
\begin{equation}
    \sigma^{abb}\left ( 0;\omega,-\omega  \right)  = -\frac{i\pi e^{3}}{2\hbar^{2}}\int \frac{dk}{8\pi^{3}} \sum_{n,m}f_{nm}
     \left (\textrm{R}_{mn}^{a,b} \right) \times \left( r_{nm}^{b} r_{mn}^{b}\right) 
     \times \delta\left ( \omega _{mn}-\omega \right).\\
    \label{shift_current_response2}
    \end{equation}
The  $\textrm{R}_{mn}^{a,b}$ is the known shift vector also known as the quantum geometric potential~\cite{PhysRevX.10.041041,WangHua2022} and is given by Equation~\ref{Rmn}
\begin{equation}
 \displaystyle 
 \textrm{R}_{mn}^{a,b} =\frac{\partial \phi_{nm}^b}{\partial k^a}+ \big(A^a_{nn}(\mathbf{k}) -A^a_{mm}(\mathbf{k}))
 \label{Rmn}
 \end{equation}
In Equation~\ref{Rmn},  the  $\phi_{nm}$ is the phase factor of the  dipole matrix elements.\\
The term $r^{b}_{nm}(\mathbf{k}) r^{b}_{mn}(\mathbf{k}) $ , in Equation~\ref{Rmn},  represents  optical transition intensity that is in agreement with previuos work~\cite{Qian2023}.  The intensity of an optical transition is related to how much light is absorbed and is also associated with the square of the dipole matrix elements. 
\section{ The band structure of the zincblende AlSb semicondutor under  tensile and compressive strain}
The band structure of the zincblende AlSb semiconductor under tensile and compressive strain. 
\begin{figure*}[htb!]
  \includegraphics[scale=0.8]{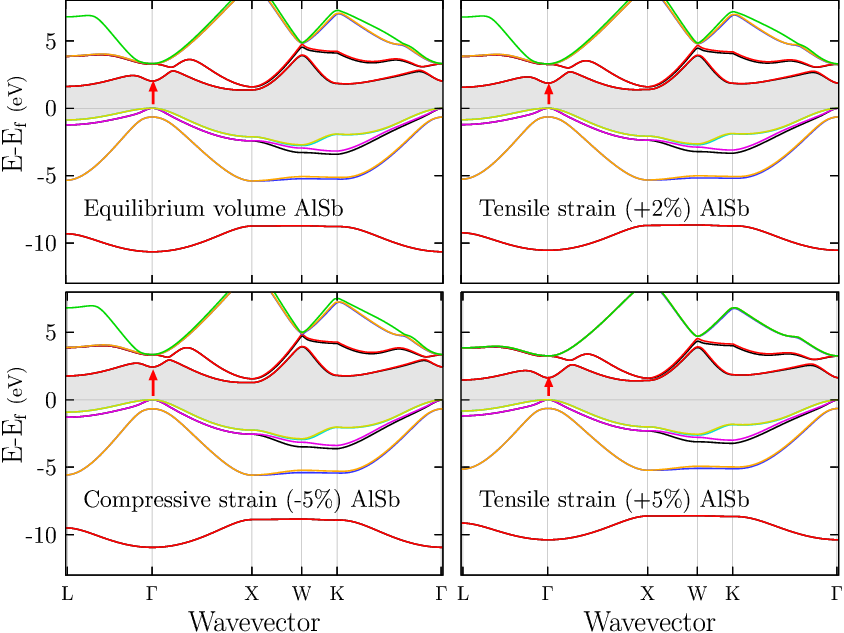}
  \caption{(Color online) Band structure of the zincblende AlSb semiconductor under tensile strain and compressive strain. The shaded grey area indicates the forbidden band energy, while the red arrows indicate the electronic band gap.}
  \label{BandsTensile}
\end{figure*}
\bibliographystyle{unsrt}
\bibliography{bibliography.bib}
\typeout{get arXiv to do 4 passes: Label(s) may have changed. Rerun}
\end{document}